\documentclass[a4paper,11pt]{article}

\usepackage{jheppub} 

\usepackage[T1]{fontenc} 

\title{}

 \author[a,b]{Marco Bochicchio}

\affiliation[a]{Scuola Normale Superiore (SNS)\\Piazza dei Cavalieri 7, Pisa, I-56100, Italy}
\affiliation[b]{INFN sez. Roma 1\\Piazzale A. Moro 2, Roma, I-00185, Italy}

\emailAdd{marco.bochicchio@roma1.infn.it}

\abstract{We prove an asymptotic structure theorem for glueball and meson propagators of any spin in large-$N$ $QCD$ and in $\mathcal{N}=1$ $SUSY$ $QCD$ with massless quarks, that determines asymptotically the residues of the poles of the propagators in terms of their anomalous dimensions and of the spectral density of the masses. The asymptotic theorem follows by the severe constraints on the propagators in large-$N$ $QCD$ with massless quarks, or in any large-$N$ confining asymptotically-free gauge theory massless in perturbation theory, that arise by perturbation theory in conjunction with the renormalization group and by the $OPE$ on the ultraviolet side.
The asymptotic theorem is inspired by a recently proposed Topological Field Theory ($TFT$) underlying large-$N$ pure $YM$, that computes sums of the scalar and of the pseudoscalar correlators satisfying the asymptotic theorem and that implies for the large-$N$ joint scalar and pseudoscalar glueball spectrum exact linearity in the masses squared. 
On the infrared side we test the prediction of the exact linearity in the $TFT$ by Meyer-Teper lattice numerical computation of the masses of the low-lying glueballs in  $SU(8)$ $YM$, finding accurate agreement.
Besides, we employ the aforementioned ultraviolet and infrared constraints in order to compare critically the scalar or pseudoscalar glueball propagators computed in the framework of the $AdS$ String/large-$N$ Gauge Theory correspondence with those of the $TFT$ underlying large-$N$ $YM$.
We find that only the $TFT$ satisfies the ultraviolet and infrared constraints. }

\usepackage{braket}
\usepackage{amsmath}
\usepackage{amssymb}
\usepackage{graphicx}
\usepackage{hyperref}
\usepackage{fancyhdr}
\newcommand{\Lambdawb}{\Lambda_{\overline{W}}}

\def\beq{\begin{equation}}
\def\eeq{\end{equation}}
\def\bea{\begin{eqnarray}}
\def\eea{\end{eqnarray}}
\def\bq{\begin{quote}}
\def\eq{\end{quote}}
\DeclareMathOperator{\Tr}{Tr}

\title{ Glueball and meson propagators of any spin in large-$N$ $QCD$}
\date{}
\begin{document}
\maketitle
%
%
%


\section{Introduction and Conclusions}

\subsection{An asymptotic structure theorem for glueball and meson propagators of any spin in large-$N$ $QCD$}

Firstly, we prove in sect.(3) an asymptotic structure theorem for glueball and meson propagators of any integer spin in 't Hooft large-$N$ limit of $QCD$ with massless quarks. In fact, the asymptotic theorem applies also to large-$N$ $\mathcal{N}=1$ $SUSY$ $QCD$ with massless quarks or to any large-$N$ confining asymptotically-free gauge theory massless to every order of perturbation theory. \par
Because of confinement we assume that the spectrum of glueball and meson masses for fixed integer spin $s$ is a discrete diverging sequence $ \{ m^{(s)}_n \}$ at the leading large-$N$ order. At the same time we assume that the spectrum $ \{ m^{(s)}_n \}$ is characterized by a smooth renormalization group ($RG$) invariant asymptotic spectral density of the masses squared $\rho_s(m^{2})$ for large masses and fixed spin, with dimension of the inverse of a mass squared, defined by:
\bea
\sum_{n=1}^{\infty} f(m^{(s)2}_n ) \sim \int_{1}^{\infty} f(m^{(s)2}_n ) dn =  \int_{ m^{(s)2}_1   }^{\infty}  f(m^2) \rho_s (m^2) dm^2
\eea
for any test function $f$. The symbol $\sim$ in this paper always means asymptotic equality in some specified sense up to perhaps a constant overall factor. \par
The asymptotic theorem reads as follows.  \par
The connected two-point Euclidean correlator of a local single-trace gauge-invariant operator $\mathcal{O}^{(s)}$, of integer spin $s$ and naive mass dimension $D$ and with anomalous dimension $\gamma_{\mathcal{O}^{(s)}}(g)$,
must factorize asymptotically for large momentum, and at the leading order in the large-$N$ limit, over the following poles and residues:
\bea \label{eq:1}
\int \langle \mathcal{O}^{(s)}(x) \mathcal{O}^{(s)}(0) \rangle_{conn}\,e^{-ip\cdot x}d^4x
\sim \sum_{n=1}^{\infty}  P^{(s)} \big(\frac{p_{\alpha}}{m^{(s)}_n}\big) \frac{m^{(s)2D-4}_n Z_n^{(s)2}  \rho_s^{-1}(m^{(s)2}_n)}{p^2+m^{(s)2}_n  } 
\eea
where $ P^{(s)} \big( \frac{p_{\alpha}}{m^{(s)}_n} \big)$ is a dimensionless polynomial in the four momentum $p_{\alpha}$ that projects on the free propagator of spin $s$ and mass $m^{(s)}_n$ and:
\bea
\gamma_{\mathcal{O}^{(s)}}(g)= - \frac{\partial \log Z^{(s)}}{\partial \log \mu}=-\gamma_{0} g^2 + O(g^4)
\eea 
with $Z_n^{(s)}$ the associated renormalization factor computed on shell, i.e. for $p^2=m^{(s)2}_n$:
\bea
Z_n^{(s)}\equiv Z^{(s)}(m^{ (s)}_n)= \exp{\int_{g (\mu)}^{g (m^{(s)}_n )} \frac{\gamma_{\mathcal{O}^{(s)}} (g)} {\beta(g)}dg}
\eea
The sum in the $RHS$ of Eq.(\ref{eq:1}) is in fact badly divergent, but the divergence is a contact term, i.e. a polynomial of finite degree in momentum. Thus the infinite sum in the $RHS$ of Eq.(\ref{eq:1}) makes sense only after subtracting the contact terms (see remark below Eq.(\ref{eq:2})). Fourier transforming Eq.(\ref{eq:1}) in the coordinate representation
projects away for $x\neq 0$ the contact terms and avoids convergence problems:
\bea \label{eq:0}
\langle \mathcal{O}^{(s)}(x) \mathcal{O}^{(s)}(0) \rangle_{conn} 
\sim \sum_{n=1}^{\infty} \frac{1}{(2 \pi)^4} \int  P^{(s)} \big(\frac{p_{\alpha}}{m^{(s)}_n}\big) \frac{m^{(s)2D-4}_n Z_n^{(s)2} \rho_s^{-1}(m^{(s)2}_n)}{p^2+m^{(s)2}_n  } \,e^{ip\cdot x}d^4p
\eea
In fact, the coordinate representation is the most convenient to get an actual proof of the asymptotic theorem, as we will see in sect.(3). \par
The physics content of the asymptotic theorem is that the residues of the poles (after analytic continuation to Minkowski space-time) are determined asymptotically by dimensional analysis, by the anomalous dimension and by the spectral density. More precisely the asymptotic behavior of the residues is fixed by the asymptotic theorem within the universal, i.e. the scheme-independent, leading and next-to-leading logarithmic accuracy.
This implies that the renormalization factors are fixed asymptotically for large $n$ to be:
\begin{equation}\label{eqn:zk_as_behav}
Z_n^{(s)2}\sim 
\Biggl[\frac{1}{\beta_0\log \frac{ m^{ (s) 2}_n }{ \Lambda^2_{QCD} }} \biggl(1-\frac{\beta_1}{\beta_0^2}\frac{\log\log \frac{ m^{ (s) 2}_n }{ \Lambda^2_{QCD} }}{\log \frac{ m^{ (s) 2}_n }{ \Lambda^2_{QCD} }}    + O(\frac{1}{\log \frac{ m^{ (s) 2}_n }{ \Lambda^2_{QCD} } } ) \biggr)\Biggr]^{\frac{\gamma_0}{\beta_0}}
\end{equation}
where $\beta_0, \beta_1,\gamma_0$ are the first and second coefficients of the beta function and the first coefficient of the anomalous dimension respectively (see for definitions subsect.(2.4) or \cite{MB}) and $\Lambda_{QCD}$
is the $QCD$ $RG$-invariant scale in some scheme. \par
The asymptotic theorem does not require any assumption
on the possible degeneracy of the spectrum for fixed spin. If there is any degeneracy it is implicit in the spectral density. We show in sect.(3) that
Eq.(\ref{eq:1}) for the propagator can be rewritten equivalently as:
\bea \label{eq:2}
\int \langle \mathcal{O}^{(s)}(x) \mathcal{O}^{(s)}(0) \rangle_{conn}\,e^{-ip\cdot x}d^4x  
\sim    P^{(s)} \big(\frac{p_{\alpha}}{p} \big)  \, p^{2D-4}   \sum_{n=1}^{\infty} \frac{Z_n^{(s)2}   \rho_s^{-1}(m^{(s)2}_n)  }{p^2+m^{(s)2}_n  }
     + \cdots
\eea
where now the sum in the $RHS$ is convergent for $\gamma'=\frac{\gamma_0}{\beta_0} > 1$. Otherwise it is divergent, but the divergence is again a contact term (see sect.(3)).
The dots in Eq.(\ref{eq:2}) represent a divergent contact term, i.e. a polynomial of finite degree in momentum\footnote{But with coefficients that are divergent in our case, because of the infinite sum.}, i.e. a distribution supported at coinciding points in the coordinate representation, and $P^{(s)} \big(\frac{p_{\alpha}}{p} \big)$ is the projector obtained substituting $-p^2$  to $m_n^2$ in  $P^{(s)} \big(\frac{p_{\alpha}}{m_n} \big)$
\footnote{We use Veltman conventions for Euclidean and Minkowski propagators of spin $s$ (see sect.(3)).}. From the proof of the asymptotic theorem in sect.(3) it follows that the divergent contact term contains at least one power of
the mass squared, i.e. two powers of $\Lambda_{QCD}$. Thus these divergent contact terms do not arise in perturbation theory.
Divergent contact terms of precisely the same kind occur in a recent computation by Zoller-Chetyrkin for the two-point glueball scalar correlator in $QCD$ by means of the standard operator product expansion ($OPE$). To mention their words \cite{chet:tensore} (p. 12): "The two-loop part is new and has a feature that did not occur in lower orders,
namely, a divergent contact term." \par
Then the proof of the asymptotic theorem reduces to showing that Eq.(\ref{eq:2})
matches asymptotically for large momentum, within the universal leading and next-to-leading logarithmic accuracy,
the $RG$-improved perturbative result \footnote{We have verified explicitly in \cite{MB} the $RG$ estimates for the operators $\Tr {F}^2$ and $\Tr {F{^*\!F}}$ on the basis of a remarkable three-loop computation by Chetyrkin et al. \cite{chetyrkin:scalar, chetyrkin:pseudoscalar} (see subsect.(1.2) and subsect.(2.4)). The earlier two-loop computation was performed in \cite{Kataev:1981gr}.} implied by the Callan-Symanzik equation (see subsect.(2.4)):
\bea \label{CS}
&& \int \langle \mathcal{O}^{(s)}(x) \mathcal{O}^{(s)}(0) \rangle_{conn}\,e^{-ip\cdot x}d^4x  \nonumber \\
&& \sim P^{(s)}\big(\frac{p_{\alpha}}{p}\big) \, p^{2D-4} Z^{(s)2}(p) \mathcal{G}_0(g(p)) \nonumber \\
&& \sim P^{(s)}\big(\frac{p_{\alpha}}{p}\big) \, p^{2D-4}    \Biggl[\frac{1}{\beta_0\log (\frac{p^2}{\Lambda^2_{QCD} } )}\biggl(1-\frac{\beta_1}{\beta_0^2}\frac{\log\log (\frac{p^2}{\Lambda^2_{QCD} } ) }{\log (\frac{p^2}{\Lambda^2_{QCD} } )}    + O(\frac{1}{\log (\frac{p^2}{\Lambda^2_{QCD} } )} ) \biggr)\Biggr]^{\frac{\gamma_0}{\beta_0}-1}
\eea
up to contact terms, and that this matching \footnote{While the asymptotic behavior of the residues in Eq.(\ref{eqn:zk_as_behav}), fixed $\gamma_0$ for the operator $\mathcal{O}$, holds for every real $\gamma'=\frac{\gamma_0}{\beta_0}$, it corresponds to the actual behavior
of the momentum representation in Eq.(\ref{CS}) for every $\gamma'$ but for $\gamma'=0,1$ (see sect.(3)).}  fixes uniquely the universal asymptotic behavior of the residues in Eq.(\ref{eq:2}). \par

Hence the meaning of the asymptotic theorem is that at large-$N$ the sum of pure poles in Eq.(\ref{eq:2}) saturates the logarithms of perturbation theory
and that the residues of the poles have a field theoretical meaning. In particular they are asymptotically proportional, apart from the power of momentum and the projector, to the square of the renormalization factor determined by the anomalous dimension divided by the spectral density, both computed on shell. \par
The asymptotic theorem has two important implications. \par
The first implication is the rather obvious observation that, given the anomalous dimension, the asymptotic spectral density can be read immediately in Eq.(\ref{eq:2}) if the residues are known for the \emph{discrete} set of poles asymptotically.
The second implication is somehow surprising. Since asymptotically we can substitute to the \emph{discrete} sum the \emph{continuous} integral weighted by the spectral density, the asymptotic propagator reads:
\bea \label{eq:3}
\int \langle \mathcal{O}^{(s)}(x) \mathcal{O}^{(s)}(0) \rangle_{conn}\,e^{-ip\cdot x}d^4x  
\sim    P^{(s)} \big(\frac{p_{\alpha}}{p} \big)  \, p^{2D-4}   \int_{ m^{(s)2}_1}^{\infty} \frac{Z^{(s)2}(m)    }{p^2+m^{2}  } dm^2
     + \cdots
\eea
with the integral representation in Eq.(\ref{eq:3}) depending only on the anomalous dimension but not on the spectral density. \par
Finally, using the Kallen-Lehmann representation (see subsect.(2.2)) we write:
\bea \label{eq:4}
&&\int \langle \mathcal{O}^{(s)}(x) \mathcal{O}^{(s)}(0) \rangle_{conn}\,e^{-ip\cdot x}d^4x  \nonumber \\
&&= \sum_{n=1}^{\infty}  P^{(s)} \big(\frac{p_{\alpha}}{m^{(s)}_n}\big) \frac{|< 0|\mathcal{O}^{(s)}(0)|p,n,s>'|^2}{p^2+m^{(s)2}_n  } \nonumber \\
&& =\sum_{n=1}^{\infty}  P^{(s)} \big(\frac{p_{\alpha}}{m^{(s)}_n}\big) \frac{m^{(s)2D-4}_n Z_n^{(s)2}  \rho_s^{-1}(m^{(s)2}_n)}{p^2+m^{(s)2}_n  }
\eea
The preceding relation between the reduced matrix elements $< 0|\mathcal{O}^{(s)}(0)|p,n,s>'$ and the renormalization factors $Z_n^{(s)}$:
\bea
|< 0|\mathcal{O}^{(s)}(0)|p,n,s>'|^2=m^{(s)2D-4}_n Z_n^{(s)2}  \rho_s^{-1}(m^{(s)2}_n)
\eea
can be regarded as a non-perturbative definition of the renormalization factors in a suitable non-perturbative scheme, in such a way that with this interpretation the asymptotic theorem holds exactly and not only asymptotically. \par
Should we know the matrix elements non-perturbatively, we would obtain also the non-perturbative contributions to the propagators due to the $OPE$. \par 
The asymptotic theorem cannot imply anything about these contributions since they are suppressed by inverse powers of momentum for large momentum. \par
The asymptotic theorem has been inspired by a computation of the anti-selfdual ($ASD$) propagator in a Topological Field Theory ($TFT$) underlying large-$N$ $YM$, that satisfies the asymptotic theorem and implies exact linearity of the joint scalar and pseudoscalar glueball spectrum, i.e. an exactly constant spectral density equal to $\Lambda_{QCD}^{-2}$ in some scheme. But the glueball propagator of the $TFT$ furnishes also the first of the non-perturbative terms in the $OPE$, that are suppressed by inverse powers of momentum, as we will see momentarily.

\subsection{Anti-selfdual glueball propagators in a Topological Field Theory underlying large-$N$ $YM$} 

Secondly, we analyze the physics implications of the anti-selfdual ($ASD$) glueball propagator computed in the aforementioned $TFT$ underlying large-$N$ pure $YM$. \par
Roughly speaking the $TFT$ describes glueball propagators in the ground state of the large-$N$ one-loop integrable sector of Ferretti-Heise-Zarembo \cite{ferretti:new_struct} (see subsect.(2.3)), that are homogeneous polynomials
of degree $L$ in the $ASD$ curvature. \par
The shortest of such operators is $ \Tr{F^-}^2(x) \equiv \sum_{\alpha \beta} \Tr {F_{\alpha \beta}^-}^2(x)$ with $F_{\alpha \beta}^-=F_{\alpha \beta} -{^*\!F}_{\alpha \beta}$ and ${^*\!}$ the Hodge dual.
In the $TFT$  \cite{boch:quasi_pbs,MB0} a non-perturbative scheme exists in which the $ASD$ glueball propagator \footnote{We use here a manifestly covariant notation
as opposed to the one in the $TFT$ \cite{boch:glueball_prop,boch:crit_points}.} is given by: 
\bea \label{eqn:top}
&& \frac{1}{2}\int \braket{\frac{g^2}{N} \Tr {F^-}^2(x) \frac{g^2}{N} \Tr{F^-}^2(0)}_{conn} \,e^{-ip\cdot x}d^4x \nonumber \\
&&= 2 \int \big(\braket{\frac{g^2}{N} \Tr {F}^2(x) \frac{g^2}{N} \Tr{F}^2(0)}_{conn} + \braket{\frac{g^2}{N} \Tr {F{^*\!F}}(x) \frac{g^2}{N} \Tr{F{^*\!F}}(0)}_{conn} \big)\,e^{-ip\cdot x}d^4x \nonumber \\
&&=\frac{1}{\pi^2}\sum_{k=1}^{\infty}\frac{(k^2 + \delta^2)g_k^4\Lambda_{\overline{W}}^6 }{p^2+k\Lambda_{\overline{W}}^2} \nonumber\\
&&=\frac{1}{\pi^2}p^4 \sum_{k=1}^{\infty}\frac{g_k^4\Lambda_{\overline{W}}^2}{p^2+k\Lambda_{\overline{W}}^2} +
 \frac{1}{\pi^2}\sum_{k=1}^{\infty}g_k^4\Lambda^2_{\overline{W}}(k\Lambda^2_{\overline{W}}-p^2)
 +\frac{1}{\pi^2}\sum_{k=1}^{\infty}\frac{ \delta^2  \, g_k^4\Lambda_{\overline{W}}^6 }{p^2+k\Lambda_{\overline{W}}^2}
\eea 
where $\Lambda_{\overline{W}}$ is the $RG$-invariant scale in the scheme in which it coincides with the mass gap, and $g_k$ is the 't Hooft coupling renormalized on shell, i.e. at $p^2=k \Lambda^2_{\overline{W}}$. The second term in the last line is a physically-irrelevant divergent sum of contact terms, i.e. a distribution supported at coinciding points in the coordinate representation. \par
It is not the aim of this paper to furnish a theoretical justification of Eq.(\ref{eqn:top}), that can be found in \cite{boch:quasi_pbs,MB0}. Additional computations can be found in \cite{boch:glueball_prop,boch:crit_points,Top}. For the purposes of this paper the reader can consider Eq.(\ref{eqn:top}) just as an ansatz that implies interesting phenomenological and theoretical consequences. In this subsection we analyze in detail these consequences. In fact, the agreement of Eq.(\ref{eqn:top}) with the $RG$, the $OPE$, and the $NSVZ$ theorem, that we discuss in this subsection, is remarkable by itself, even without the theoretical justification in \cite{boch:quasi_pbs,MB0}. \par
Eq.(\ref{eqn:top}) contains a new term proportional to $\delta^2$ that in a previous computation \cite{boch:glueball_prop,boch:crit_points,MB0} was set to zero by a Wick-ordering prescription, necessary to cancel, as in ordinary $YM$ perturbation theory of composite operators, certain infinite contributions in the $TFT$. This computation will be reported elsewhere. \par
We show momentarily that Novikov-Shifman-Vainshtein-Zakharov ($NSVZ$) low-energy theorem (see subsect.(2.5)) fixes instead the residual finite part, arising after the arbitrary subtraction due to Wick-ordering, so that $\delta$ does not actually vanish. \par
We have checked by direct computation in \cite{MB} in collaboration with S. Muscinelli that the $ASD$ propagator of the $TFT$ satisfies asymptotically \footnote{In \cite{MB} we have set $\delta=0$.}:
\bea \label{7}
 &&\frac{1}{\pi^2}\sum_{k=1}^{\infty}\frac{(k^2 + \delta^2)g_k^4\Lambda_{\overline{W}}^6 }{p^2+k\Lambda_{\overline{W}}^2}  \nonumber \\
&&  \sim  \frac{ p^{4} }{\pi^2 \beta_0}  \Biggl[\frac{1}{\beta_0\log (\frac{p^2}{\Lambda^2_{\overline{W}} } )}\biggl(1-\frac{\beta_1}{\beta_0^2}\frac{\log\log (\frac{p^2}{\Lambda^2_{\overline{W}} } ) }{\log (\frac{p^2}{\Lambda^2_{ \overline{W}} } )}    + O(\frac{1}{\log (\frac{p^2}{\Lambda^2_{\overline{W}} } )} ) \biggr)\Biggr]
\eea
up to contact terms, according to the asymptotic theorem of this paper and to the fact that the first coefficient of the anomalous dimension of $\Tr F^{-2}$ is $\gamma_0=2 \beta_0$ \cite{MB}.
In fact, the inspiration for the proof of the asymptotic theorem came from the computation \cite{boch:glueball_prop,boch:crit_points} in the $TFT$ and from the detailed $RG$ estimates in \cite{MB} (see subsect.(2.4)). \par
But Eq.(\ref{eqn:top}) contains a finer information than the asymptotic theorem. \par
Indeed, on the $UV$ side Eq.(\ref{eqn:top}) reproduces the first two coefficient functions in the $RG$-improved $OPE$ of the $ASD$ propagator (see subsect.(2.4)):
\bea \label{OP}
&& \frac{1}{2} \int \braket{\frac{g^2}{N}\Tr{F^-}^2(x) \frac{g^2}{N}\Tr{F^-}^2(0)}_{conn} \,e^{-ip\cdot x}d^4x \nonumber \\
&& \sim  C_0(p^2)  +C_1(p^2) <\frac{g^2}{N}\Tr{F^-}^2(0)>+...
\eea
and not only the first coefficient, i.e. the perturbative contribution
implied by the asymptotic theorem. $C_0(p^2)$ is the perturbative coefficient function displayed in Eq.(\ref{7}):
\bea \label{C}
C_0(p^2)
\sim \frac{ p^{4} }{\pi^2 \beta_0}   \Biggl[\frac{1}{\beta_0\log (\frac{p^2}{\Lambda^2_{\overline{W}} } )}\biggl(1-\frac{\beta_1}{\beta_0^2}\frac{\log\log (\frac{p^2}{\Lambda^2_{\overline{W}} } ) }{\log (\frac{p^2}{\Lambda^2_{ \overline{W}} } )}    + O(\frac{1}{\log (\frac{p^2}{\Lambda^2_{\overline{W}} } )} ) \biggr)\Biggr]
\eea
and $C_1(p^2)$ is fixed by the general principles of the $RG$ and by the Callan-Symanzik equation to satisfy asymptotically (see subsect.(2.4)):
\bea
  C_1(p^2) \sim \frac{ 1 }{\pi^2 \beta_0} \Biggl[\frac{1}{\beta_0\log (\frac{p^2}{\Lambda^2_{\overline{W}} } )}\biggl(1-\frac{\beta_1}{\beta_0^2}\frac{\log\log (\frac{p^2}{\Lambda^2_{\overline{W}} } ) }{\log (\frac{p^2}{\Lambda^2_{ \overline{W}} } )}    + O(\frac{1}{\log (\frac{p^2}{\Lambda^2_{\overline{W}} } )} ) \biggr)\Biggr]
\eea
Indeed, it corresponds in Eq.(\ref{eqn:naive_rg1}) to the case $D=4, D_1=4, \gamma_0(\mathcal{O}_D)=2 \beta_0$ and, since the glueball condensate is $RG$ invariant,$\gamma_0(\mathcal{O}_{D_1})=0$. 
The scalar contribution to $C_1(p^2)$ arising from the scalar propagator in the second line of Eq.(\ref{eqn:top}) has been computed recently at two-loop order by Zoller-Chetyrkin \cite{chet:tensore} in the $\overline{MS}$ scheme.
Disregarding momentarily the contact terms in Eq.(\ref{eqn:top}), the same estimates that enter the proof of the asymptotic theorem in sect.(3) or in \cite{MB} imply:
\bea \label{9}
 \frac{1}{\pi^2}\sum_{k=1}^{\infty}\frac{ \delta^2 g_k^4\Lambda_{\overline{W}}^6 }{p^2+k\Lambda_{\overline{W}}^2} 
&&  \sim  \Lambda_{\overline{W}}^4 \frac{ \delta^2}{\pi^2 \beta_0}   \Biggl[\frac{1}{\beta_0\log (\frac{p^2}{\Lambda^2_{\overline{W}} } )}\biggl(1-\frac{\beta_1}{\beta_0^2}\frac{\log\log (\frac{p^2}{\Lambda^2_{\overline{W}} } ) }{\log (\frac{p^2}{\Lambda^2_{ \overline{W}} } )}    + O(\frac{1}{\log (\frac{p^2}{\Lambda^2_{\overline{W}} } )} ) \biggr)\Biggr] \nonumber \\
&&\sim  \delta^2  \Lambda_{\overline{W}}^4 C_1(p^2)
\eea
Thus the $TFT$ is in perfect agreement with the constraint arising by the perturbative $OPE$ and the $RG$ also for the second coefficient function in the $OPE$. \par
Besides, the glueball condensate  $<\frac{g^2}{N}tr{F^-}^2(0)>$ is non-vanishing in the $TFT$ \cite{boch:crit_points,MB0}, as opposed to perturbation theory. Its value in the $TFT$ is proportional to a suitable power of the $RG$-invariant scale.
Let us call this scale $\Lambda_{GC}$:
\bea
<\frac{g^2}{N}\Tr{F^-}^2(0)>=\Lambda_{GC}^4
\eea
Moreover, the zero-momentum divergent sum of contact terms in Eq.(\ref{eqn:top}) mixes with $C_1(p^2) <\frac{g^2}{N}tr{F^-}^2(0)>$ in the $OPE$ implicitly determined by the $ASD$ propagator of the $TFT$, in such a way that $C_1(p^2)$ in the $TFT$ has a zero-momentum quadratically-divergent part. \par
Remarkably, a similarly divergent contact term at zero momentum occurs in the recent perturbative computation by Zoller-Chertyrkin \cite{chet:tensore} of the part of the second coefficient $C_1(p^2)$ that arises from the scalar propagator contributing to the $ASD$ correlator, and it is an obstruction to implementing the $NSVZ$ theorem (see subsect.(2.5)):
\bea \label{8}
&&\int \braket{\frac{g^2}{N}\Tr F^2(x) \frac{g^2}{N}\Tr{F^-}^2(0)}_{conn} \,d^4x 
=\frac{4}{\beta_0}\braket{\frac{g^2}{N}\Tr{F^-}^2(0)}
\eea
in perturbation theory, since in perturbation theory the subtraction of the infinite zero-momentum contact term in the $LHS$ leaves a finite ambiguity in the zero-momentum correlator, that affects the $RHS$ of Eq.(\ref{8}). \par
To mention again Zoller-Chertyrkin words \cite{chet:tensore}: "The two-loop part is new and has a feature that did not occur in lower orders,
namely, a divergent contact term. Its appearance clearly demonstrates that non-logarithmic
perturbative contributions to $C_1$ are not well defined in $QCD$, a fact seemingly ignored by the
the $QCD$ sum rules practitioners." \par
The aforementioned infinite ambiguity is resolved in the $TFT$ because of the unambiguous non-perturbative
separation between the contact terms and the physical terms that carry the pole singularities (in Minkowski space-time) in Eq.(\ref{eqn:top}),
and the subsequent  subtraction of the quadratically-divergent sum of contact terms displayed in Eq.(\ref{eqn:top}).\par
Indeed, in the $TFT$ the $NSVZ$ theorem reads \footnote{This follows from the identity $\Tr F^2(x)=\frac{1}{2} \Tr F^{-2}(x)+\Tr(F{^*\!F})$ and by the fact that the term $\int d^4x \Tr(F{^*\!F})$ is irrelevant in the $TFT$ \cite{boch:crit_points}.}(see subsect.(2.5)):
\bea \label{10}
&&\int \braket{\frac{g^2}{N} \Tr{F^-}^2(x) \frac{g^2}{N}\Tr{F^-}^2(0)}_{conn} \,d^4x 
=\frac{8}{\beta_0}\braket{\frac{g^2}{N}\Tr{F^-}^2(x)}
\eea
After subtracting the contact terms it combines with Eq.(\ref{eqn:top}) to give:
\bea
\frac{2}{\pi^2}\sum_{k=1}^{\infty}\frac{ \delta^2  \, g_k^4\Lambda_{\overline{W}}^6 }{k\Lambda_{\overline{W}}^2}= \frac{8}{\beta_0} \Lambda_{GC}^4
\eea
where the convergent series in the LHS arises as the restriction to zero momentum of the third term in the last line in Eq.(\ref{eqn:top}). 
Thus the $NSVZ$ theorem fixes $\delta$ and, as a consequence, the normalization of the first non-trivial coefficient function in the $OPE$ of the $TFT$. \par
On both the infrared ($IR$) and the ultraviolet ($UV)$ side Eq.(\ref{eqn:top}) is not only an asymptotic formula but implies exact linearity in the square of the masses of the joint scalar and pseudoscalar spectrum in the large-$N$ limit of $YM$ all the way down to the low-lying glueball states. \par
This is a strong statement that could be easily falsified.  \par
Indeed, on the infrared side it implies that the ratio of the masses of the two lowest-scalar (or pseudoscalar) glueball states is $\sqrt2=1.4142 \cdots$.
As we discuss in subsect.(1.5), in the lattice computation that is presently closer to the continuum limit \footnote{This means on the presently larger lattice with the smaller value of the $YM$ coupling.} for $SU(8)$ $YM$, Meyer-Teper \cite{Me1,Me2} found for the mass ratios of the lowest scalar and pseudoscalar states, $r_s=\frac{m_{0^{++*}}}{m_{0^{++}}}$ and $r_{ps}=\frac{m_{0^{-+}}}{m_{0^{++}}}$, $r_s=r_{ps}=1.42(11)$ in accurate agreement with the $TFT$. In  subsect.(1.5) we compare the predictions of the $TFT$ also with the lattice computations of Lucini-Teper-Wenger \cite{L1} and of Lucini-Rago-Rinaldi \cite{L2}. \par
In addition, on the infrared side it is needed a non-perturbative definition of the beta function in order for Eq.(\ref{eqn:top}) to make sense, since for the low-lying glueballs $g_k$ must be evaluated at scales on the order of 
$\Lambda_{\overline{W}}$ and this is a scale close, if not coinciding, to the one where the perturbative Landau infrared singularity of the running coupling occurs. \par
The $TFT$ provides such a non-perturbative scheme for the beta function for which no Landau infrared singularity of the coupling occurs \cite{boch:quasi_pbs}. \par
The functions $g(\frac{p}{\Lambdawb})$ and $Z(\frac{p}{\Lambdawb})$ are the solutions of the differential equations \cite{boch:quasi_pbs}:
\begin{align} \label{eqn:eq_def_gk}
\frac{\partial g}{\partial \log p}
&=\frac{-\beta_0 g^3+\frac{1}{(4\pi)^2}g^3\frac{\partial \log Z}{\partial \log p}}{1-\frac{4}{(4\pi)^2}g^2} \nonumber \\
\frac{\partial\log Z}{\partial\log p}
&=2\gamma_0 g^2 +\cdots \nonumber \\
\gamma_{0}&=\frac{1}{(4\pi)^2}\frac{5}{3}
\end{align}
with $p=\sqrt{p^2}$.
The definitions of $g_k$ and $Z_k$ are:
\begin{align}
&g_k=g(\sqrt k)\\
&Z_k=Z( \sqrt k)\label{01}
\end{align}
In \cite{boch:quasi_pbs} it is shown that Eq.(\ref{eqn:eq_def_gk}) reproduces the correct universal one-loop and two-loop coefficients of the perturbative $\beta$ function of pure $YM$. Indeed, we get:
\begin{align}\label{eqn:matching_beta_pert}
\frac{\partial g}{\partial \log p}
&=\frac{-\beta_0 g^3+\frac{2\gamma_0 }{(4\pi)^2}g^5}{1-\frac{4}{(4\pi)^2}g^2}+\cdots \nonumber \\
&=-\beta_0 g^3+\frac{2\gamma_0}{(4\pi)^2}  g^5-\frac{4\beta_0}{(4\pi)^2}g^5+\cdots \nonumber\\
&=-\beta_0 g^3-\beta_1 g^5+\cdots
\end{align}
with:
\begin{align}
&\beta_0=\frac{1}{(4\pi)^2}\frac{11}{3}\\
&\beta_1=\frac{1}{(4\pi)^4}\frac{34}{3}
\end{align}
Besides, in the $TFT$ the glueball propagators for the operators $\mathcal{O}_{2L}$ in the ground state of Ferretti-Heise-Zarembo \cite{ ferretti:new_struct} can be computed \cite{boch:glueball_prop} asymptotically for large $L$ \footnote{Again we have set  $\delta=0$ in Eq.(\ref{eqn:formula}).}. These operators have mass dimension $D=2L$ and are homogeneous polynomials of degree $L$ in the $ASD$ curvature $F^-$ \cite{ferretti:new_struct} (see subsect.(2.3)):
\begin{align}\label{eqn:formula}
&\int \braket{\mathcal{O}_{2L}(x)\mathcal{O}_{2L}(0)}_{conn}  \,e^{-ip\cdot x}d^4x  
\sim \sum_{k=1}^{\infty}\frac{k^{2L-2} Z_k^{-L}\Lambda_{\overline{W}}^2 \Lambda_{\overline{W}}^{4L-4}}{p^2+k\Lambda_{\overline{W}}^2} 
\end{align}
Ferretti-Heise-Zarembo have computed the one-loop anomalous dimension of $\mathcal{O}_{2L}$ for large $L$ \cite{ferretti:new_struct}:
\begin{align}
\gamma_{0 (\mathcal{O}_{2L})}= \frac{1}{(4\pi)^2}\frac{5}{3} L+O(\frac{1}{L})
\end{align}
The one-loop anomalous dimension computed within the $TFT$ Eqs.(\ref{eqn:eq_def_gk}-\ref{01}-\ref{eqn:formula})
agrees with Ferretti-Heise-Zarembo computation asymptotically for large $L$ and exactly for the $L=2$ ground state, that is the $ASD$ operator that occurs in Eq.(\ref{eqn:top}), for which $\gamma_{0 (\mathcal{O}_{4})}=2 \beta_0$ exactly. \par
As a consequence the asymptotic theorem of this paper is satisfied asymptotically for large-$L$ by the large-$L$ $ASD$ correlators of the $TFT$ as well, as it has been checked by direct computation in \cite{MB}.

\subsection{The $AdS$/Gauge Theory correspondence versus the Topological Field Theory} 

Thirdly, we compare the proposal for the glueball propagators of the $TFT$ with the widely known proposals for the large-$N$ glueball propagators of a vast class of confining $QCD$-like theories, including pure $YM$, $QCD$ and $SUSY$ gauge theories, based on the $AdS$/Large-$N$ Gauge Theory correspondence. \par
In the framework of the $AdS$/Large-$N$ Gauge Theory correspondence \cite{Mal} we examine Witten supergravity background \cite{Witten}, that has been proposed to describe large-$N$ $QCD$,
and Klebanov-Strassler supergravity background \cite{KS1,KS2}, that has been proposed to describe large-$N$ cascading $\mathcal{N}$ $=1$ $SUSY$ gauge theories. They belong to the so called top-down approach, that means that
they are essentially deductions from first principles in the framework of the $AdS$/Large-$N$ Gauge Theory correspondence. Therefore, they are very rigid and lead to sharp predictions for the glueball spectrum and the glueball propagators. \par
Also the $TFT$ underlying large-$N$ $YM$ is meant to be a deduction from fundamental principles \cite{boch:quasi_pbs,MB0} and therefore it is very rigid and leads to a sharp prediction for the joint scalar and pseudoscalar glueball spectrum and propagator as well. \par
We examine also Polchinski-Strassler model \cite{PS,PS2} or Hard-Wall model and the Soft-Wall model \cite{Softwall}. They belong to the bottom-up approach in the framework of the $AdS$/Large-$N$ Gauge Theory correspondence, that means that they are 
meant to be models that aim to incorporate some features of large-$N$ $QCD$ rather than deductions from fundamental principles. Therefore, they are less rigid and consequently their predictions are not as sharp
as in the previous cases. For example, the spectrum of the Hard-Wall model depends on the choice of boundary conditions at the wall \cite{Brower2}. The spectrum of the Soft-Wall model \cite{Softwall} depends on the ad hoc choice of the dilaton potential, that purposely
is chosen in such a way to imply exact linearity of the square of glueball and meson masses, as opposed to the spectrum of the Hard-Wall model \cite{Brower2}, of Witten model \cite{Brower1} and of Klebanov-Strassler background \cite{KS1,KS2}, that are asymptotically quadratic in the square of the glueball masses. \par
All these different proposals can be tested both in the infrared and in the ultraviolet. \par
The infrared test is by numerical results in lattice gauge theories. \par
The ultraviolet test is by first principles. Indeed, as we pointed out in the previous subsections, the structure of the glueball propagators is severely constrained by the perturbative $RG$, as the asymptotic theorem of this paper shows, and by the $OPE$. Another test by first principles is by the low-energy theorems of $NSVZ$, that we have discussed in the framework of the $TFT$. A short review of the theoretical background behind these ideas is
reported in sect.(2).
\par
We should add at this stage that all the proposals that are meant to describe large-$N$ $YM$ or large-$N$ $QCD$, i.e. Witten background, the Hard-Wall model, the Soft-Wall model and the $TFT$, sharply disagree \footnote{The only common feature is the gross picture of the existence of the mass gap and of an infinite tower of massive glueballs.}
among themselves both about the $IR$ low-energy spectrum and about the $UV$.\par

\subsection{The ultraviolet test}

We have submitted the aforementioned proposals to a stringent test in the $UV$ for the asymptotics of the scalar and/or pseudscalar glueball propagator, that coincides up to an overall constant with $C_0$ in Eq.(\ref{C}) \cite{MB}, after which only the $TFT$ has survived. Indeed, in the framework of the $AdS$ String/Large-$N$ Gauge Theory correspondence all the glueball propagators, for which we could find presently an explicit computation in the literature, behave as $p^4 \log^n(\frac{p^2}{\mu^2})$, with
$n=1$ for the Hard- and Soft-Wall models \cite{forkel,italiani,forkel:holograms,forkel:ads_qcd} and $n=3$ for Klebanov-Strassler background \cite{krasnitz:cascading2,krasnitz:cascading}, in contradiction with the universal $RG$ estimate \cite{MB} for $C_0$ Eq.(\ref{C}). \par 
 Klebanov-Strassler background deserves a further separate examination. \par
There is no infrared test for it, since no lattice computation is available for supersymmetric gauge theories. \par
Moreover, it has not passed the ultraviolet test for the scalar glueball propagator \cite{MB}, despite it is able to reproduce even in the supergravity approximation
the correct $NSVZ$ asymptotically-free $\beta$ function of the large-$N$ cascading $\mathcal{N}$ $=1$ $SUSY$ gauge theories. Since this is puzzling, we suggest here a possible explanation. \par
Indeed, in $\mathcal{N}$ $=1$ $SUSY$ $YM$, the final end of the cascade, it there exists a phase strongly coupled in the $UV$ foreseen by Kogan-Shifman \cite{KSh}.
This phase is described by the very same large-$N$ $NSVZ$ $\beta$ function:
\bea
\frac{\partial g}{\partial \log \Lambda}=-\frac{\frac{3}{\left(4\pi\right)^2} g^3}
{1- \frac{2}{\left(4\pi\right)^2}  g^2 }
\eea
since the $IR$ fixed point of the $RG$ flow $g^2 = \frac{\left(4\pi\right)^2}{2}$ is attractive both for $g^2 \leq \frac{\left(4\pi\right)^2}{2}$, the asymptotically-free phase weakly-coupled in the $UV$, and for $g^2 \geq \frac{\left(4\pi\right)^2}{2}$, the strongly-coupled phase in the $UV$.  Therefore, Kogan-Shifman argue \cite{KSh} that there exists a strongly-coupled phase in the $UV$, admitting a continuum limit, described by the strong-coupling branch of the same $NSVZ$ beta function, whose weak coupling branch describes the asymptotically-free phase. \par In \emph{both} cases the $RG$ flow stops at $g^2 = \frac{\left(4\pi\right)^2}{2}$, so that the running coupling \emph{never} diverges in the $IR$. In particular the $RG$
flow is \emph{not} connected to $g^2=\infty$ in the $IR$. \par 
However, the $RG$ flow is connected to $g^2=\infty$ in the $UV$ of the non-asymptotically-free phase. \par
In fact, it is natural to identify the aformentioned strongly-coupled phase in the $UV$ with Klebanov-Strassler background, since the effective coupling of the corresponding scalar glueball propagator grows in the $UV$ as $\log^3(\frac{p^2}{\mu^2})$ \cite{krasnitz:cascading2,krasnitz:cascading}
instead of decreasing as $\frac{1}{\log(\frac{p^2}{\mu^2})}$, as the universal estimate for $C_0$ in the asymptotically-free phase would require \cite{MB}.\par
Thus we are led to conclude that even in the most favorable situation, when the exact $\beta$ function is reproduced on the string side of the correspondence, the $AdS$ String/large-$N$ Gauge Theory correspondence in its present strong coupling incarnation describes in a neighborhood of $g^2=\infty$ the aforementioned \emph{strongly-coupled} phase in the $UV$, whose existence is implied by the supersymmetric $NSVZ$ $\beta$ function, \emph{not} the asymptotically-free phase. \par
But lattice gauge theory computations in $YM$ (or in $QCD$ in 't Hooft large-$N$ limit) show that the aforementioned strongly-coupled phase in the $UV$, admitting a continuum limit, does not exist 
in pure non-supersymmetric $YM$.

\subsection{The infrared test}

We are interested in the large-$N$ limit, therefore we look for lattice results that have been computed for the largest gauge group possible. \par
We should mention that comparisons of this kind have been already presented in the past years by many groups, using the lattice results for $SU(3)$ as benchmark.
But in recent years lattice results for larger gauge groups
up to $SU(8)$ have become available, as opposed to the earlier important $SU(3)$ results (for an updated review of large-$N$ lattice $QCD$ see \cite{ML}). \par
Since for all the approaches proposed in the literature the computations are supposed to hold in the large-$N$ limit, there is not much point in looking at lattice result for $SU(3)$
once lattice results for higher rank $SU(N)$ groups have become available. If $SU(3)$ is sufficiently close to $SU(N)$, as some evidence from the numerical lattice results seems to approximately indicate, the $SU(N)$ result will be a good description of both. If not, the theoretical predictions that we want to test are meant for large-$N$ $SU(N)$
and not for $SU(3)$. Therefore $SU(8)$ is presently the most suitable choice in this framework.  \par
Thus we compare in some detail the predictions for the low-lying glueball masses, scalar, pseudoscalar and spin 2, with the three lattice numerical computations for $SU(8)$, discussing also the lattice numerical uncertainty. \par
There are presently three lattice computations, in chronological order, by
Lucini-Teper-Wenger \cite{L1}, by Meyer-Teper \cite{Me1,Me2} and by Lucini-Rago-Rinaldi \cite{L2} for the mass ratios, $r_s=\frac{m_{0^{++*}}}{m_{0^{++}}}$, $r_{ps}=\frac{m_{0^{-+}}}{m_{0^{++}}}$ and $r_2= \frac{m_{2^{++}}}{m_{0^{++}}}$ in $SU(8)$ $YM$.
They are remarkably in agreement when compared on the same lattice and for close values of the $YM$ coupling. Since Lucini-Teper-Wenger and Lucini-Rago-Rinaldi essentially agree at quantitative level, we discuss in detail for
simplicity only the most recent computation, i.e. Lucini-Rago-Rinaldi, that we compare with Meyer-Teper. \par
However, Meyer-Teper perform the computation also for one smaller value of the $YM$ coupling and a larger lattice and perhaps a different variational basis, in order to be as close as possible to the continuum limit. \par
As a consequence there is about a $20\%$ difference in their final results:
For Meyer-Teper $r_s=r_{ps}=1.42(11)$ and for Lucini-Rago-Rinaldi: $r_s=1.79(08)$, $r_{ps}=1.78(08)$. Yet both computations show degeneracy of the first excited scalar with the first pseudoscalar mass. In addition, the mass ratio of the lowest spin-$2$ glueball to the lowest scalar is for Meyer-Teper $r_2= \frac{m_{2^{++}}}{m_{0^{++}}} = 1.40  $ while for Lucini-Rago-Rinaldi $r_2= \frac{m_{2^{++}}}{m_{0^{++}}} = 1.70 $. \par 
A possible interpretation is that new states arise for smaller coupling corresponding to the ratios $r_s=r_{ps}=1.42(11)$ of Meyer-Teper \footnote{We would like to thank Biagio Lucini for suggesting this interpretation.}. \par
Of course the previous observation implies that Meyer-Teper is closer to the continuum limit, but their result should be taken with a grain of salt because Meyer-Teper computation is presently the only one for such a smaller coupling. \par
Indeed, the previous Lucini-Teper-Wenger computation $r_s \sim 1.83$ is in agreement with Lucini-Rago-Rinaldi. Yet it has been suggested \footnote{Biagio Lucini, private communication.} that $r_s=1.79(08)$, $r_{ps}=1.78(08)$
is quite close to the prediction of the $TFT$ for the next-excited glueballs,  $r_{s}=r_{ps}=\sqrt3=1.7320 \cdots$, if it is assumed that that Lucini-Rago-Rinaldi see only the next-excited glueballs for some reason linked to the choice of the variational basis and/or
the value of the $YM$ coupling. This should be clarified by future computations. \par
The theoretical predictions are as follows. \par
In the $TFT$, $r_{s}=r_{ps}=\sqrt2=1.4142 \cdots$ in accurate agreement with Meyer-Teper. For the second scalar or pseudoscalar excited state the $TFT$ predicts  $r_{s}=r_{ps}=\sqrt3=1.7320 \cdots$, quite close to  Lucini-Rago-Rinaldi values, if we assume that they do not see the lower state of Meyer-Teper. \par
In Witten model $r_s=1.5860$, $r_{ps}=1.2031$, $r_2=1$. These numbers are obtained from \cite{Brower1} according to the standard identification (see also \cite{Mal} for the numerical values of $r_s$ and $r_{ps}$) of the dilaton on the string side as the dual of $\Tr F^2$ on the gauge side \footnote{
On the contrary the standard identification is not employed in \cite{Brower1}. In fact, in \cite{Brower1} it is shown that on the string side there is another scalar state with mass lower than the dilaton. But this lowest-mass scalar couples to a field on the string side that has no correspondent on the gauge side. In particular, according to the non-standard identification, the mass gap would not arise by states that couple to $\Tr F^2$, a statement that we do not believe. For this non-standard choice $r_s=1.7388$, $r_{ps}=2.092$, $r_2=1.7388$. Indeed, subsequently in \cite{Brower2} it is employed the standard identification.}. \par
In the Hard-Wall model (Polchinski-Strassler) for $Dirichlet$ boundary conditions \cite{Braga1,Braga2} $r_s=1.64$, $r_2=1.48$, for \emph{Neumann} boundary conditions \cite{Braga2} $r_s=1.83$, $r_2=1.56$, while for other \emph{different} boundary conditions for \emph{different} states \cite{Brower2} $r_s=2.19$, $r_{ps}=1.25$, $r_2=1.25$. \par
In the Soft-Wall model \cite{forkel,italiani,forkel:holograms,forkel:ads_qcd} $r_s=\sqrt\frac{3}{2}=1.2247 \cdots$  \par
Thus the $TFT$ agrees sharply with Meyer-Teper.  \par
Witten model is inconsistent with Lucini-Rago-Rinaldi and barely compatible with Meyer-Teper for $r_s$ or $r_{ps}$ taken separately, but it is in contrast with their apparent degeneracy implied by the lattice result of both groups. On the contrary it predicts $r_2=1$,
i.e. that the lowest-mass spin-$2$ glueball is exactly degenerate with the lowest-mass scalar, that is sharply in contradiction with all the lattice computations, not only Meyer-Teper. \par
The Soft-Wall model is barely compatible with Meyer-Teper and inconsistent with Lucini-Rago-Rinaldi. \par
The Hard-Wall model is very sensitive to boundary conditions and thus the question is as to whether it can \emph{fit} the lattice data, rather than predict anything.
Yet none of the choices of boundary conditions gives an accurate prediction for $r_s$ but in one case: For \emph{Neumann} bounday conditions and assuming that Lucini-Rago-Rinaldi see the first excited state and Meyer-Teper 
computation is not correct.  In addition, in the Hard-Wall model as in Witten model, $r_2=1$ \cite{Brower2} unless rather arbitrarily the boundary conditions
for the scalar and the spin-$2$ glueball are chosen to be different. \par
Our conclusion is that Meyer-Teper lattice computation clearly favors the $TFT$ in the infrared and disfavors all the other models considered. \par
Besides, it is desirable that Meyer-Teper computation be confirmed and extended by other groups \footnote{Biagio Lucini communicated to us that there is an ongoing computation by Lucini-Rago-Rinaldi.}. \par

\subsection{Conclusions}

We have proved an asymptotic structure theorem for glueball and meson propagators of any integer spin in large-$N$ $QCD$ that fixes asymptotically the residues of the poles in terms of the anomalous dimension and of the spectral
density \footnote{After this paper was posted in the arXiv
we have been informed of \cite{A1,A2} where, for the meson propagators of the scalar and of the vector current in $QCD$,
the scaling of the residues with the meson masses are analyzed
assuming an asymptotically linear spectrum and employing a different technique
based on dispersion relations and on the explicit perturbative computation.
The leading and next-to-leading asymptotic results of \cite{A1,A2} for the residues of the meson propagator of the vector and of the scalar current agree perfectly
with the asymptotic theorem of this paper as special cases.}. \par
The asymptotic theorem was inspired by a $TFT$ underlying large-$N$ $YM$. \par
The $ASD$ glueball propagator of the $TFT$ satisfies the constraints that follow by the perturbative
renormalization group, i.e. the asymptotic theorem, and by the first non-perturbative term in the $OPE$ as well. However, the $TFT$ does not contain a complete set of condensates of operators
in the $OPE$. This is not surprising since the $TFT$ is supposed to describe by construction only the ground state of Ferretti-Heise-Zarembo one-loop integrable sector of large-$N$ $YM$. \par
Moreover, none of the scalar or pseudoscalar propagators based on the $AdS$ String/ large-$N$ Gauge Theory correspondence presently computed in the literature, as opposed to the $TFT$, satisfies any of the constraints that arise
by the renormalization group and by the $OPE$ in the $UV$. \par
In particular, somehow surprisingly, Klebanov-Strassler background does not reproduce the universal $UV$ asymptotics of $\mathcal{N}$ $=1$ $SUSY$ $YM$, despite it reproduces the
correct beta function. We suggest as explanation that it describes the phase not asymptotically free but strongly coupled in the ultraviolet
foreseen by Kogan-Shifman on the basis of the structure of the $NSVZ$ beta function. \par
On the infrared side the $TFT$ agrees accurately with Meyer-Teper lattice computation, the mass spectra based on the  presently proposed versions of the $AdS$ String/Gauge Theory correspondence do not. \par
We conclude that the glueball propagator of the $TFT$ is definitely favored by first principles in the $UV$, and presently by lattice data in the $IR$, with respect to the glueball propagators of the $AdS$ String/Gauge Theory correspondence in its present strong coupling incarnation.

\section{A short review of the large-$N$ limit of $QCD$}

\subsection{'t Hooft large-$N$ limit}

The  $SU(N)$ pure $YM$ theory is defined by the partition function:
\begin{equation}\label{eqn:Z1}
Z=\int \delta A \, e^{-\frac{1}{2g_{YM}^2}\int \sum_{\alpha\beta} \Tr \bigl(F_{\alpha\beta}^2\bigr)d^4x}
\end{equation}
Introducing 't Hooft coupling constant $g$ \cite{'t hooft:large_n}:
\begin{equation}
g^2=g^2_{YM}N
\end{equation}
the partition function reads:
\begin{equation}
Z=\int \delta A\, e^{-\frac{N}{2g^2}\int \sum_{\alpha\beta} \Tr \bigl(F_{\alpha\beta}^2\bigr)   d^4x}
\end{equation} 
According to 't Hooft \cite{'t hooft:large_n} the large-$N$ limit is defined with $g$ fixed when $N\rightarrow \infty$. 
The normalization of the action in Eq.(\ref{eqn:Z1}) corresponds to choosing the gauge field $A_\alpha=A^a_\alpha t^a $ with the generators $t^a$ valued in the fundamental representation of the Lie algebra, normalized as:
\begin{equation}
\Tr\, (t^a t^b)=\frac{1}{2}\delta^{ab}
\end{equation}
In Eq.(\ref{eqn:Z1}) $F_{\alpha \beta}$ is defined by:
\begin{equation}\label{eqn:F_wilsonian}
F_{\alpha\beta}(x)=\partial_\alpha A_\beta-\partial_\beta A_\alpha + i[A_\alpha,A_\beta]
\end{equation}
We refer to the normalization of the action in Eq.(\ref{eqn:Z1}) as the Wilsonian normalization. 
However, perturbation theory is formulated with the canonical normalization (employed in subsect.(1.2)), obtained rescaling the field $A_\alpha $ in Eq.(\ref{eqn:Z1}) by the coupling constant $g_{YM}=\frac{g}{\sqrt{N}}$:
\begin{align}
A_\alpha (x)\rightarrow g_{YM}A^c_\alpha (x)
\end{align}
in such a way that in the action the kinetic term becomes independent on $g$:
\begin{equation}
\frac{1}{2}\int \sum_{\alpha \beta} \Tr(F_{\alpha \beta} ^2(A^c)) (x) d^4 x
\end{equation}
where:
\begin{align}\label{eqn:F_canonical}
F_{\alpha \beta}(A^c)= \partial_\beta A^c_\alpha  - \partial_\alpha  A^c_\beta +ig_{YM}[A^c_\alpha  , A^c_\beta]
\end{align}
In 't Hooft large-$N$ limit \cite{'t hooft:large_n} $r$-point connected correlators of single-trace local operators with the Wilsonian normalization scale as $N^{2-r}$. It follows that at the leading $\frac{1}{N}$ order multi-point correlators of local gauge invariant operators factorize:
\begin{align} \label{eqn:ordine_1_N}
&\braket{\mathcal{O}_1(x_1)\mathcal{O}_2(x_2)\cdots \mathcal{O}_n(x_n)}
\nonumber\\
&=\braket{\mathcal{O}_1(x_1)}\braket{\mathcal{O}_2(x_2)}\cdots \braket{\mathcal{O}_n(x_n)}
+O(1)
\end{align}
Indeed, according to Eq.(\ref{eqn:ordine_1_N}), the one-point correlators are of order of $N$, while the connected two-point correlators are of order of 1.
The connected three-point correlators are of order of $\frac{1}{N}$ and so on. Therefore, only one-point condensates survive at leading order and two-point connected correlators survive at next-to-leading order. Hence the interaction vanishes in the large-$N$ limit at the leading order for connected correlators, since it is associated to the three- and multi-point connected correlators. \par

\subsection{Kallen-Lehmann representation of two-point correlators}

Because of confinement and the mass gap and the vanishing of the interaction at the leading large-$N$ order, it is believed \cite{migdal:multicolor} that the two-point connected Euclidean correlators of local gauge invariant single-trace  scalar operators $\mathcal{O}^{(0)}(x)$ in the pure glue sector of large-$N$ $QCD$:
\bea
G^{(2)}_{conn}(p)=\int \langle \mathcal{O}^{(0)}(x)\mathcal{O}^{(0)}(0) \rangle_{conn}e^{- ip\cdot x} d^4x = \int^{\infty}_0 \frac{\mathcal{R}(m)}{p^2+m^2} dm^2 
\eea
are an infinite sum of propagators of massive free fields, i.e. the spectral distribution $\mathcal{R}(m)$ in the Kallen-Lehmann representation is saturated by massive free one-particle states only, the glueballs \cite{migdal:multicolor,{polyakov:gauge}}. In the scalar or pseudoscalar case:
\bea
G^{(2)}_{conn}(p)&& = \sum_{n=1}^{\infty}   \frac{|< 0|\mathcal{O}^{(0)}(0)|p,n>|^2}{p^2+m^{(0)2}_n  } \nonumber \\
&& = \sum_n\frac{\mathcal{R}_{n}}{p^2+m^{(0) 2}_{n}}
\eea
The generalization to any integer spin \cite{migdal:multicolor}, that includes also gauge-invariant fermion bilinears in the large-$N$ 't Hooft limit of $QCD$, is:
\bea \label{uv}
\int \langle \mathcal{O}^{(s)}(x)\mathcal{O}^{(s)}(0) \rangle_{conn}e^{-ip\cdot x} d^4x
= \sum_{n=1}^{\infty}  P^{(s)} \big(\frac{p_{\alpha}}{m^{(s)}_n}\big) \frac{|< 0|\mathcal{O}^{(s)}(0)|p,n,s>'|^2}{p^2+m^{(s)2}_n  } 
\eea
In \cite{migdal:multicolor} Migdal pointed out that the sum in Eq.(\ref{uv}) must be infinite, otherwise it cannot be asymptotic to the perturbative result. \par
The asymptotic theorem of subsect.(1.1) and sect.(3) is in fact a quantitative refinement of this statement. \par
The reduced matrix elements $< 0|\mathcal{O}^{(s)}(0)|p,n,s>'$  are expressed in terms of the polarization vectors $e^{(s)}_j(\frac{p_\alpha}{m})$ and of the matrix elements $< 0|\mathcal{O}^{(s)}(0)|p,n,s,j>$ of the operator $\mathcal{O}^{(s)}$ between the vacuum and one-particle states $|p,n,s,j>$:
\bea
< 0|\mathcal{O}^{(s)}(0)|p,n,s,j>= e^{(s)}_j(\frac{p_\alpha}{m}) < 0|\mathcal{O}^{(s)}(0)|p,n,s>'
\eea
The polarization vectors define the projectors that enter the spin-$s$ propagators:
\bea
\sum_j e^{(s)}_j(\frac{p_\alpha}{m})  \overline{e^{(s)}_j(\frac{p_\alpha}{m})}= P^{(s)} \big(\frac{p_{\alpha}}{m}\big)
\eea
The free propagators for $s=1,2$ were worked out in \cite{Velt} (see the end of sect.(3) for explicit formulae). The generalization to any integer or half-integer spin can be found in \cite{Francia1,Francia2}.

\subsection{The large-$N$ integrable sector of Ferretti-Heise-Zarembo}

In the 't Hooft large-$N$ limit of $QCD$ there is a special sector of the theory discovered by Ferretti-Heise-Zarembo \cite{ferretti:new_struct}, that is integrable at one-loop for the anomalous dimensions. \par
The  pure glue subsector of the integrable sector is composed by local single-trace gauge invariant operators built by the anti-selfdual ($ASD$) or the selfdual ($SD$) part of the curvature $F_{\alpha \beta}$ and their covariant derivatives \cite{ferretti:new_struct}.
They are defined by:
\begin{align}
F^-_{\alpha \beta}&= F_{\alpha \beta}- {^*\!F}_{\alpha \beta}\nonumber\\
F^+_{\alpha \beta}&= F_{\alpha \beta}+ {^*\!F}_{\alpha \beta}
\end{align}
where:
\begin{equation}
{^*\!F}_{\alpha \beta}=\frac{1}{2}\epsilon_{\alpha\beta\gamma \delta}F^{\gamma\delta}
\end{equation}
Therefore, the operators in the subsector described above have the form:
\begin{equation}
\mathcal{O}(x)=\Tr(D_{\mu_1}\cdots D_{\mu_n}F^-_{\alpha_1\beta_1}D_{\nu_1}\cdots D_{\nu_m}F^-_{\alpha_2 \beta_2}\cdots\cdots
D_{\rho_1}\cdots D_{\rho_l}F^-_{\alpha_L\beta_L})(x) 
\end{equation}
with any possible contraction of the indices.
Here $L$ is the number of $F^-$ in the operator $\mathcal{O}$.
This sector is integrable at one loop in the large-$N$ limit \cite{ferretti:new_struct}. 
The anomalous dimensions of these operators 
can computed at one loop as the eigenvalues of the Hamiltonian of a closed spin chain. The construction extends to chiral fermion bilinear operators of massless quarks and to
an open spin chain \cite{ferretti:new_struct}. \par
The ground state of the Hamiltonian spin chain by definition corresponds to the operators with the most negative anomalous dimensions. For any fixed $L$ the ground state of the closed chain turns out to be built by operators that contain only $F^-_{\alpha\beta}$ and that have indices contracted to obtain a scalar in a peculiar way determined by the anti-ferromagnetic ground state of the spin chain:
\begin{equation}\label{eqn:op_interessanti}
\mathcal{O}_{2L}(x)=\Tr(\,\underbrace{F^-_{\alpha_1\beta_1}\cdots F^-_{\alpha_L\beta_L}}_{{}\text{Certain scalar contractions}}\,)(x)
\end{equation}
with dimension in energy $D=2L$. In the spin chain each $F^-_{\alpha_i\beta_i}$ corresponds to a site, therefore $L$ corresponds to the length of the chain.
Hence the large $L$ limit corresponds to the thermodynamic limit, i.e the infinite length limit.
In \cite{ferretti:new_struct} it was computed the large-$N$ one-loop anomalous dimension of the ground state of the spin chain of length $L$, using the Bethe ansatz in the thermodynamic limit:
\begin{align}\label{eqn:intro_dim_anomala_L}
\gamma_{\mathcal{O}_{2L}}(g)&=-\gamma_0\, L\, g^2+O(\frac{1}{L}) \nonumber\\
\gamma_0&=\frac{5}{3}\frac{1}{(4\pi)^2}
\end{align}
For $L=2$ the operator in the ground state is $\Tr {F^-}^2$ and its one-loop anomalous dimension is exactly (see also \cite{MB}):
\begin{align}
\gamma_{\mathcal{O}_4}(g)&=-2\beta_0 \, g^2 + \cdots \nonumber \\
\beta_0&=\frac{11}{3}\frac{1}{(4\pi)^2}
\end{align} 
The $\mathcal{O}_4$ correlator reduces in Euclidean space-time to the sum of the scalar $\mathcal{O}_S=\Tr F^2$ and pseudoscalar correlator $\mathcal{O}_P=\Tr F {^*\!F} $:
\bea
\frac{1}{2} \langle \mathcal{O}_4(x)\mathcal{O}_4(0) \rangle_{conn}=  2 \langle \mathcal{O}_S(x)\mathcal{O}_S(0) \rangle_{conn}+  2 \langle \mathcal{O}_P(x)\mathcal{O}_P(0) \rangle_{conn}
\eea
 
 \subsection{Renormalization group and $OPE$}
 
The structure of the two-point correlators of (scalar) local gauge invariant operators in $QCD$ with massless quarks or in any asymptotically free gauge theory with no perturbative mass scale is severely constrained \cite{migdal:multicolor} by perturbation theory in conjunction with the renormalization group \cite{MB} and by the operator product expansion ($OPE$) \cite{migdal:multicolor}:
\bea
\int \langle \mathcal{O}_D(x)\mathcal{O}_D(0) \rangle_{conn}e^{-ip\cdot x}d^4x= C_0(p^2)+C_1(p^2) <\mathcal{O}_{D_1}(0)>+ \cdots
\eea
Assuming multiplicative renormalizability of the operator  $\mathcal{O}_D$,
the coefficient functions $C_0, C_1, \cdots$ in the $OPE$ satisfy the Callan-Symanzik equations (see for example \cite{ZI}):
\begin{equation}\label{eqn:RG_eq_p}
\left(p_{\alpha}\frac{\partial}{\partial p_{\alpha}}-\beta(g)\frac{\partial}{\partial g}-2(D-2+\gamma_{\mathcal{O}_D}(g))\right)C_0(p^2)=0
\end{equation}
and:
\begin{equation}\label{eqn:RG_eq_p}
\left(p_{\alpha}\frac{\partial}{\partial p_{\alpha}}    -\beta(g)\frac{\partial}{\partial g}-(2D-D_1-4+2\gamma_{\mathcal{O}_D}(g)- \gamma_{\mathcal{O}_{D_1}}(g)  )\right)C_1(p^2)=0
\end{equation}
The solution for $C_0$ is \cite{MB}:
\begin{equation}\label{eqn:pert_general_behavior}
C_0(p^2)= p^{2D-4}\,\mathcal{G}_{0}(g(p))\, Z^2_{\mathcal{O}_D}(\frac{p}{\mu},g(p))
\end{equation}
and:
\begin{equation}
C_1(p^2)= p^{2D-D_1-4}\,\mathcal{G}_{1}(g(p))\, Z^2_{\mathcal{O}_D}(\frac{p}{\mu},g(p)) Z^{-1}_{\mathcal{O}_{D_1}}(\frac{p}{\mu},g(p))
\end{equation}
with: 
\bea
\gamma_{\mathcal{O}_D}(g)= - \frac{\partial \log Z_{\mathcal{O}_D}}{\partial \log \mu}=-\gamma_{0}(\mathcal{O}_D) g^2 +\cdots
\eea 
and:
\bea
\beta(g)= \frac{\partial g}{\partial \log \mu}=-\beta_{0} g^3 - \beta_1 g^5 +\cdots
\eea 
The power of $p$ is implied by dimensional analysis, $\mathcal{G}$ is a dimensionless function that depends only on the running coupling $g(p)$ and $Z$ is the contribution from the anomalous dimension.\par
Since the correlator of composite operators is conformal at the lowest non-trivial order in perturbation theory, the perturbative estimate for $\mathcal{G}_0,\mathcal{G}_1$ is \cite{MB}:
\begin{equation}
\mathcal{G}(g(p)) \sim {\log \frac{p^2}{ \Lambda^2_{QCD}} }\sim \frac{1}{g^2(p)}
\end{equation}
Indeed, $\int p^{2D-4} \log \frac{p^2}{\mu^2} e^{ ipx} d^4p \sim \frac{1}{x^{2D}}$ is conformal in the coordinate representation for $D$ integer, $D \ge 3$ (see Appendix A of \cite{MB}). \par 
Collecting the previous results, we get the naive scheme-independent universal large-momentum asymptotic estimate for $C_0$ \cite{MB}:
\begin{align}\label{eqn:naive_rg}
&C_0(p^2)\sim p^{2D-4}
g(p)^{\frac{2\gamma_0(\mathcal{O}_D)  }{\beta_0}-2}
\end{align}
and analogously for $C_1$:
\begin{align}\label{eqn:naive_rg1}
&C_1(p^2)\sim p^{2D-D_1-4}
g(p)^{\frac{2\gamma_0(\mathcal{O}_D)  - \gamma_0(\mathcal{O}_{D_1}) }{\beta_0}-2}
\end{align}
In fact, these estimates are naive because the correlator of $\mathcal{O}_D$ in the momentum representation is not multiplicatively renormalizable because of the presence of contact terms in perturbation theory. \par
Thus the naive $RG$-estimates may hold only after subtracting the contact terms. 
The strategy to check them is as follows. \par
In the coordinate representation \cite{chetyrkin:TF} no contact term arises for $x\neq 0$.
If:
\bea
 \langle \mathcal{O}_D(x)\mathcal{O}_D(0) \rangle_{conn}= C_0(x^2)+C_1(x^2) <\mathcal{O}_{D_1}(0)>+ \cdots
\eea
the coefficient functions $C_0, C_1, \cdots$ in the $OPE$ satisfy the Callan-Symanzik equations (see for example \cite{ZI}):
\begin{equation}\label{eqn:RG_eq_p}
\left(x_{\alpha}\frac{\partial}{\partial x_{\alpha}}+\beta(g)\frac{\partial}{\partial g}+2(D+\gamma_{\mathcal{O}_D}(g))\right)C_0(x^2)=0
\end{equation}
and:
\begin{equation}\label{eqn:RG_eq_p}
\left(x_{\alpha}\frac{\partial}{\partial x_{\alpha}}    +\beta(g)\frac{\partial}{\partial g}+(2D-D_1+2\gamma_{\mathcal{O}_D}(g)- \gamma_{\mathcal{O}_{D_1}}(g)  )\right)C_1(x^2)=0
\end{equation}
The solutions are:
\begin{equation}\label{eqn:pert_general_behavior}
C_0(x^2)= \frac{1}{x^{2D}}\,\mathcal{G}_{0}(g(x))\, Z^2_{\mathcal{O}_D}(x \mu,g(x))
\end{equation}
and:
\begin{equation}
C_1(x^2)= \frac{1}{x^{2D-D_1}}\,\mathcal{G}_{1}(g(x))\, Z^2_{\mathcal{O}_D}(x \mu,g(x)) Z^{-1}_{\mathcal{O}_{D_1}}(x \mu,g(x))
\end{equation}
with $x=\sqrt{x^2}$.
Since the correlator is conformal at the lowest non-trivial order in perturbation theory, the perturbative estimate for $\mathcal{G}(g(x))$ is \cite{MB}:
\begin{equation}
\mathcal{G}(g(x)) \sim 1 + O(g^2(x))
\end{equation}
Collecting the previous results, we get the actual small-distance scheme-independent universal asymptotic behavior:
\begin{align}
&C_0(x^2)\sim \frac{1}{x^{2D}} \,
g(x)^{\frac{2\gamma_0(\mathcal{O}_D)  }{\beta_0}}
\end{align}
and:
\begin{align}\label{eqn:naive_rg}
&C_1(x^2)\sim \frac{1}{x^{2D-D_1}} \,
g(x)^{\frac{2\gamma_0(\mathcal{O}_D)  - \gamma_0(\mathcal{O}_{D_1}) }{\beta_0}}
\end{align}
Thus, in order to get the correct $RG$ estimates in the momentum representation, we should first compute the Fourier transform of the $RG$-improved result in the coordinate representation. But in general the Fourier transform does not exist because of the local singularity in $x=0$.
Nevertheless, as a byproduct of the proof of the asymptotic theorem, we show in sect.(3) how to obtain explicit results for the large-momentum asymptotics of the Fourier transform, \emph{after} the subtraction of the contact terms.
It turns out that the naive $RG$ estimate in the momentum representation for $C_0$ is in fact correct, but in the two cases $\gamma'=0,1$ with $\gamma'=\frac{\gamma_0}{\beta_0}$, that need only
a slight refinement discussed in sect.(3). Entirely similar results hold for $C_1$.
For the case $\gamma'=0$ the asymptotic estimate in the momentum representation is simply $C_0(p^2) \sim p^{2D-4} \log \frac{p^2}{\mu^2}$, that corresponds to a correlator asymptotically conformal in the $UV$ (see Appendix A of \cite{MB}).

\subsection{$NSVZ$ low-energy theorems in $QCD$}

We adapt to the large-$N$ limit the derivation of the low-energy theorem  in \cite{1,2}, for a scalar operator $\mathcal{O}_D$ with dimension in energy $D$ and anomalous dimension $\gamma_{\mathcal{O}_D}$. \par
Actually, in order to make contact with the $TFT$ of subsect.(1.2), we specialize to the operators $\mathcal{O}_{2L}$, that occur as the ground state of the Hamiltonian spin chain in the integrable sector of Ferretti-Heise-Zarembo. While in intermediate steps we consider the large-$L$ limit, the actual formulation of the $NSVZ$ theorem depends only on the dimension $D$ of the operator. \par
We present the derivation for an operator with generic anomalous dimension, while originally $NSVZ$ considered only the $RG$-invariant case, i.e. zero anomalous dimension. \par
We start by the definition:
\begin{equation}\label{eqn:LE_def}
\braket{\frac{1}{N} \Tr \mathcal{O}_D}=\frac{\int \frac{1}{N} \Tr \mathcal{O}_D(0) e^{-\frac{N}{2g^2}\int \Tr F^2(x)d^4x }}
{\int e^{-\frac{N}{2g^2}\int \Tr F^2(x)d^4x }}
\end{equation}
and we assume that there exists a non-perturbative scheme in which:
\begin{equation*}
\braket{\frac{1}{N} \Tr\mathcal{O}_D}=\Lambda_{YM}^D Z_{\mathcal{O}_D}
\end{equation*}
In addition for large-$L$, in the ground state of Ferretti-Heise-Zarembo:
\begin{equation*}
 Z_{\mathcal{O}_{2L}}=Z^{L+\mathit{O}(\frac{1}{L})}
\end{equation*}
for some $Z$.
We derive both members of Eq.(\ref{eqn:LE_def})
with respect to $-\frac{1}{g^2}$.
Therefore, for large $L$:
\begin{equation*}
\frac{\partial\braket{\frac{1}{N} \Tr \mathcal{O}_{2L}}}{\partial(-\frac{1}{g^2})}\sim
2L\,\Lambda_{YM}^{2L-1}\,\frac{\partial \Lambda_{YM}}{\partial(-\frac{1}{g^2})}\,Z^{L}+
L Z^{L-1}\Lambda_{YM}^{2L}\frac{\partial Z}{\partial(-\frac{1}{g^2})}
\end{equation*}
To compute $\frac{\partial \Lambda_{YM}}{\partial(-\frac{1}{g^2})}$ we use the definition of $\Lambda_{YM}$:
\begin{equation*}
\left(\frac{\partial}{\partial \log \Lambda}+\beta(g)\frac{\partial}{\partial g}\right)\Lambda_{YM}=0
\end{equation*}
so that:
\begin{equation*}
\frac{\partial \Lambda_{YM}}{\partial(-\frac{1}{g^2})}=\frac{g^3}{2}\frac{\partial\Lambda_{YM}}{\partial g}=
-\frac{g^3}{2\beta(g)}\frac{\partial \Lambda_{YM}}{\partial \log\mu}=
-\frac{g^3}{2\beta(g)}\Lambda_{YM}
\end{equation*}
The last identity follows from the relation:
\begin{equation*}
\Lambda_{YM}=\Lambda f(g)= e^{\log \Lambda} f(g)
\end{equation*}
for some function $f(g)$.
To compute $\frac{\partial Z}{\partial(-\frac{1}{g^2})}$ we use its definition:
\begin{align*}
Z=e^{\int_{g(\mu)}^{g(\Lambda)}\frac{\gamma(g')}{\beta(g')}dg'} \nonumber\\
\Rightarrow \frac{\partial Z}{\partial(-\frac{1}{g^2})}=
\frac{g^3}{2\beta(g)}Z \gamma(g)
\end{align*}
On the other hand, deriving the RHS of Eq.(\ref{eqn:LE_def}) we get:
\begin{equation*}
\frac{\partial\braket{\frac{1}{N}\Tr\mathcal{O}_{2L}}}{\partial(-\frac{1}{g^2})}=
\frac{1}{2}\int \braket{\Tr\mathcal{O}_{2L}(0) \Tr F^2(x)}_{conn}d^4x
\end{equation*}
and:
\begin{align*}
- \frac{g^3}{\beta(g)} D(1-\frac{\gamma(g)}{2}) \braket{\frac{1}{N} \Tr\mathcal{O}_D}
=  \int \braket{\Tr\mathcal{O}_D(0)  \Tr F^2(x)}_{conn}d^4x
\end{align*}
with the Wilsonian normalization of the action. Finally, taking the limit $\Lambda \rightarrow \infty$ we get the $NSVZ$ low-energy theorem with the Wilsonian normalization
of the action:
\begin{align*}
 \frac{D}{\beta_0}  \braket{\frac{1}{N} \Tr\mathcal{O}_D}
=  \int \braket{\Tr\mathcal{O}_D(0)  \Tr F^2(x)}_{conn}d^4x
\end{align*}
\section{The asymptotic structure theorem for glueball and meson propagators of any spin in large-$QCD$}

Firstly, we prove the asymptotic theorem for scalar or pseudoscalar propagators. \par
We define the asymptotic spectral density as follows.
For any test function $f$ we assume that the spectral sum can be approximated asymptotically by an integral, keeping the leading term in the Euler-MacLaurin formula \cite{migdal:meromorphization}:
\bea
\sum_{n=1}^{\infty} f(m^{(s)2}_n ) \sim \int_{1}^{\infty} f(m^{(s)2}_n ) dn
\eea
Then by definition the asymptotic spectral density satisfies:
\bea
\frac{d n}{d m^{(s)2}}=\rho_s(m^2)
\eea
i.e. :
\bea
 \int_{1}^{\infty} f(m^{(s)2}_n ) dn =  \int_{ m^{(s)2}_1   }^{\infty}  f(m^2) \rho_s (m^2) dm^2
 \eea
We write an ansatz for the large-$N$
two-point Euclidean correlator of a local gauge-invariant scalar or pseudoscalar operator $\mathcal{O}$ of naive dimension in energy $D$ and with anomalous dimension $\gamma_{\mathcal{O}}(g)$:
\bea \label{p}
\int \langle \mathcal{O}(x) \mathcal{O}(0) \rangle_{conn}\,e^{-ip\cdot x}d^4x  
=  \sum_{n=1}^{\infty} \frac{ R_n m^{2D-4}_n \rho^{-1}(m_n^2) } {p^2+m^2_n }
\eea
This ansatz in not restrictive and follows only by dimensional analysis to the extent the dimensionless pure numbers $R_n$ are unspecified yet.
However, the specific form of the ansatz is the most convenient for our aims. \par
We now distinguish two cases, $D$ even and $D$ odd. For local gauge-invariant composite operators in $QCD$ the lowest non-trivial operator with $D$ even occurs for $D=4$ in the pure glue sector, while the lowest $D$ odd occurs for $D=3$ in the sector containing fermion bilinears.
For $D$ even using the identity:
\bea \label{b}
m^{2D-4}_n =((m^{2}_n +p^2)(m^{2}_n-p^2)+p^4)^{\frac{D}{2}-1}
\eea
we get:
\bea \label{b0}
\int \langle \mathcal{O}(x) \mathcal{O}(0) \rangle_{conn}\,e^{-ip\cdot x}d^4x  
= p^{2D-4} \sum_{n=1}^{\infty} \frac{R_n  \rho^{-1}(m_n^2) }{p^2+m^2_n } + \cdots
\eea
where the dots represent contact terms, i.e. distributions whose Fourier transform is supported at $x=0$, that are physically irrelevant and that therefore can be safely discarded. 
The contact terms arise because, for $D$ even and $\frac{D}{2}-1$ positive, in Eq.(\ref{b}) in addition to the term $p^{2D-4}$ at least one term involving the factor of $m^{2}_n +p^2$, that cancels the denominator, always occurs. \par
For $D$ odd we use instead the identity:
\bea \label{b1}
m_n^{2D-4} = m_n^2 m^{2(D-1)-4}_n =
(p^2+m_n^2-p^2) ((m^{2}_n +p^2)(m^{2}_n-p^2)+p^4)^{\frac{D-1}{2}-1}
\eea
from which we get a similar result but with opposite sign:
\bea \label{b2}
\int \langle \mathcal{O}(x) \mathcal{O}(0) \rangle_{conn}\,e^{-ip\cdot x}d^4x  
= - p^{2D-4} \sum_{n=1}^{\infty} \frac{R_n  \rho^{-1}(m_n^2) }{p^2+m^2_n } + \cdots
\eea
It is also clear from Eq.(\ref{b}) and Eq.(\ref{b0}) that the sum of the contact terms is divergent, but nevertheless the entire sum, and not just the individual terms, is a polynomial of \emph{finite} degree in momentum. Later in this subsection we pass to the coordinate representation, where contact terms do no arise at all for $x\ne 0$. A fact that confirms their physical irrelevance.
\par
Now we substitute to the sum the integral using the Euler-McLaurin formula:
 \begin{equation}
\sum_{k=k_1}^{\infty}G_k(p)=
\int_{k_1}^{\infty}G_k(p)dk - \sum_{j=1}^{\infty}\frac{B_j}{j!}\left[\partial_k^{j-1}G_k(p)\right]_{k=k_1}
\end{equation}
We disregard the terms involving the Bernoulli numbers since in our case they are suppressed by inverse powers of momentum. Thus the infinite sum reads asymptotically:
\bea
&& \sum_{n=1}^{\infty} \frac{R_n  \rho^{-1}(m_n^2) }{p^2+m^2_n }  \nonumber \\
&&\sim\int_{1}^{\infty} \frac{R_n  \rho^{-1}(m_n^2) }{p^2+m^2_n } dn \nonumber \\
&&= \int_{m^2_1}^{\infty} \frac{R(m)  \rho^{-1}(m^2) }{p^2+m^2 } \rho(m^2) dm^2 \nonumber \\
&&= \int_{m^2_1}^{\infty} \frac{R(m) }{p^2+m^2 } dm^2
\eea
Now we compare Eq.(\ref{p}) with perturbation theory.
Assuming asymptotic freedom the non-perturbative propagator has to match at large momentum, up to contact terms,
the large momentum $RG$-improved perturbative result obtained solving the Callan-Symanzik equation, that assuming naively multiplicative renormalizability of the operator $\mathcal{O}$ reads (see subsect.(2.4)):
\bea
\int \langle \mathcal{O}(x) \mathcal{O}(0) \rangle_{conn}\,e^{-ip\cdot x}d^4x  
\sim p^{2D-4} Z_{\mathcal{O}}^{2}(p) \mathcal{G}_{0}(g(p))
\eea
This assumption is too naive because of the occurrence of contact terms also in perturbation theory. However, we prove
later, employing the coordinate representation of the propagator, that after subtracting the contact terms in the momentum representation the naive $RG$-estimate is in fact correct but in the special cases $\gamma'=0,1$ with $\gamma'=\frac{\gamma_0}{\beta_0}$. \par
The only unknown function is $\mathcal{G}_{0}(g(p))$ that is supposed to be a $RG$-invariant function of the running coupling only.
$\mathcal{G}_{0}(g(p))$ is fixed for a composite operator at the lowest non-trivial order by the condition that the two-point correlator be exactly conformal in the $UV$ in the coordinate representation.   \par
Hence we must have asymptotically for large $p$:
\bea
&& \int_{m^2_1}^{\infty} \frac{R(m) }{p^2+m^2 } dm^2 = Z_{\mathcal{O}}^{2}(p) \mathcal{G}_{0}(g(p))
\eea
up perhaps to an overall sign. It is convenient first to compactify the $dm^2$ integration and then to remove the cutoff $\Lambda$. For large $\Lambda$ and for large $p<< \Lambda$:
\bea
&& \int_{m^2_1}^{\Lambda^2} \frac{R(m) }{p^2+m^2 } dm^2 = Z_{\mathcal{O}}^{2}(p) \mathcal{G}_{0}(g(p))
\eea 
This is an integral equation of Fredholm type for which, by the Fredholm alternative, a solution exists if and only if it is unique. We find first explicitly a solution for large $\Lambda$, then we show how it extends to $\Lambda=\infty$.
It is convenient to introduce the dimensionless variables $\nu=\frac{p^2}{\Lambda^2_{QCD}}$, $k=\frac{m^2}{\Lambda^2_{QCD}}$
and $K=\frac{\Lambda^2}{\Lambda^2_{QCD}}$. We get:
\bea
&& \int_{k_1}^{K} \frac{R(\sqrt k) }{\nu+k} dk= Z_{\mathcal{O}}^{2}(\sqrt \nu) \mathcal{G}_{0}(g(\sqrt \nu))
\eea 
and explicitly (see subsect.(2.4)), keeping only the asymptotic universal part:
\bea \label{12}
&& \int_{k_1}^{K} \frac{R(\sqrt k) }{\nu+k} dk= 
\Biggl(\frac{1}{\beta_0\log \nu}\biggl(1-\frac{\beta_1}{\beta_0^2}\frac{\log\log\nu}{\log\nu}\biggr)\Biggr)^{\frac{\gamma_0}{\beta_0}-1}
\eea
We show now that the solution is:
\bea \label{11}
R(\sqrt k) \sim Z^2(\sqrt k) \sim \Biggl(\frac{1}{\beta_0\log\frac{k}{c}}\biggl(1-\frac{\beta_1}{\beta_0^2}\frac{\log\log\frac{k}{c}}{\log\frac{k}{c}} + O(\frac{1}{\log k}) \biggr)  \Biggr)^{\frac{\gamma_0}{\beta_0}}
\eea
with asymptotic accuracy for large $k$ in the sense determined by the term $O(\frac{1}{\log k})$, i.e. within the universal leading and next-to-leading logarithmic accuracy, as remarked in sect.(1). The constant $c$ is related to the scheme dependence, but the universal part is actually $c$ independent. 
The proof of existence of the solution is by direct computation. The necessary integrals have been already computed in \cite{MB}.  We substitute the ansatz in Eq.(\ref{11}) into Eq.(\ref{12}). We distinguish two cases: either $\gamma'>1$ or otherwise. For $\gamma'>1$
the integral in Eq.(\ref{12}) is convergent, in such a way that the integration domain can be extended to $\infty$. Otherwise the integral is divergent, but the divergence is a contact term. Therefore, after subtracting the contact term, the solution can be extended to $\infty$. Following \cite{MB} firstly we change variables in  the $LHS$ of Eq.(\ref{12}) from $k$ to $k+\nu$:
\begin{align} \label{14}
I_c^{2}(\nu)&=\int_1^{\infty}\beta_0^{-\gamma'}\left(\frac{1}{\log(\frac{k}{c})}\left(1-\frac{\beta_1}{\beta_0^2}\frac{\log\log(\frac{k}{c})}{\log(\frac{k}{c})}\right)\right)^{\gamma'}\frac{dk}{k+\nu}\nonumber\\
&=\beta_0^{-\gamma'}\int_{1+\nu}^{\infty}\left(\frac{1}{\log(\frac{k-\nu}{c})}\left(1-\frac{\beta_1}{\beta_0^2}\frac{\log\log(\frac{k-\nu}{c})}{\log(\frac{k-\nu}{c})}\right)\right)^{\gamma'}\frac{dk}{k}\nonumber\\
&\sim \beta_0^{-\gamma'}\int_{1+\nu}^{\infty}\left[\log(\frac{k-\nu}{c})\right]^{-\gamma'}
\left(1-\gamma'\frac{\beta_1}{\beta_0^2}\frac{\log\log(\frac{k-\nu}{c})}{\log(\frac{k-\nu}{c})}\right)\frac{dk}{k}\nonumber\\
&\sim  \beta_0^{-\gamma'}\int_{1+\nu}^{\infty}\left[\log(\frac{k-\nu}{c})\right]^{-\gamma'}\frac{dk}{k} 
-\gamma'\frac{\beta_1}{\beta_0^2}\beta_0^{-\gamma'}\int_{1+\nu}^{+\infty}\left[\log(\frac{k-\nu}{c})\right]^{-\gamma'-1}\log\log(\frac{k-\nu}{c})
\frac{dk}{k}
\end{align}
For the first integral in the last line we get:
\begin{align}\label{eqn: int1}
&\int_{1+\nu}^\infty\frac{1}{k}[\beta_0\log(\frac{k}{c})]^{-\gamma'}
\biggl[1+\frac{\log(1-\frac{\nu}{k})}{\log(\frac{k}{c})}\biggr]^{-\gamma'}dk\nonumber\\
&\sim \int_{1+\nu}^\infty\frac{1}{k}[\beta_0\log(\frac{k}{c})]^{-\gamma'} 
\biggl[1+\gamma'\frac{\nu}{k \log(\frac{k}{c})}\biggr]dk\nonumber\\
&=\int_{1+\nu}^\infty\frac{1}{k}[\beta_0\log(\frac{k}{c})]^{-\gamma'}dk+
\gamma' \nu\int_{1+\nu}^\infty\frac{1}{k^2}\beta_0^{-\gamma'}[\log(\frac{k}{c})]^{-\gamma'-1}dk 
\end{align}
From the first integral it follows the leading asymptotic behavior \cite{boch:glueball_prop} provided $\gamma' \neq 1$:
\begin{equation}\label{eqn:sol_int_leading}
\int_{1+\nu}^\infty\frac{1}{k}[\beta_0\log(\frac{k}{c})]^{-\gamma'}dk=
\frac{1}{\gamma'-1} \beta_0^{-\gamma'}\left[\log\left(\frac{1+\nu}{c}\right)\right]^{-\gamma'+1}
\end{equation}
For $\gamma' = 0$ there is nothing to add. It corresponds to the asymptotically conformal case in the $UV$. If $\gamma' \neq 0$ we
add the second contribution. We evaluate it at the leading order by changing variables and integrating by parts:
\begin{align}\label{eqn:int_next-to-leading}
&\gamma'\frac{\beta_1}{\beta_0^2}\beta_0^{-\gamma'}\int_{1+\nu}^{+\infty}\left[\log(\frac{k-\nu}{c})\right]^{-\gamma'-1}\log\log(\frac{k-\nu}{c})
\frac{dk}{k}\nonumber\\
&\sim  \gamma'\frac{\beta_1}{\beta_0^2}\beta_0^{-\gamma'}\int_{1+\nu}^{+\infty}\left[\log(\frac{k}{c})\right]^{-\gamma'-1}\log\log(\frac{k}{c})
\frac{dk}{k}\nonumber\\
&=\gamma'\frac{\beta_1}{\beta_0^2}\beta_0^{-\gamma'}\int_{\log\frac{1+\nu}{c}}^{+\infty}t^{-\gamma'-1}\log(t)dt\nonumber\\
&=\gamma'\frac{\beta_1}{\beta_0^2}\beta_0^{-\gamma'}\left[\frac{1}{\gamma'}\left(\log(\frac{1+\nu}{c})\right)^{-\gamma'}\log\log(\frac{1+\nu}{c})+
\frac{1}{\gamma'^2}\left(\log(\frac{1+\nu}{c})\right)^{-\gamma'}\right]
\end{align}
The second term in brackets in the last line is subleading with respect to the first one. 
Collecting Eq.(\ref{eqn:int_next-to-leading}) and Eq.(\ref{eqn:sol_int_leading}) we get for $I_c^{2}(\nu)$:
\begin{align}\label{eqn:esp_ntl}
&\beta_0^{-\gamma'}\int_1^{\infty}\left(\frac{1}{\log(\frac{k}{c})}\left(1-\frac{\beta_1}{\beta_0^2}\frac{\log\log(\frac{k}{c})}{\log(\frac{k}{c})}\right)\right)^{\gamma'}\frac{dk}{k+\nu}\nonumber\\
& \sim \frac{1}{\gamma'-1}\beta_0^{-\gamma'}\left(\log\frac{1+\nu}{c}\right)^{-\gamma'+1}-\frac{\beta_1}{\beta_0^2}\beta_0^{-\gamma'}\left(\log(\frac{1+\nu}{c})\right)^{-\gamma'}\log\log(\frac{1+\nu}{c})\nonumber\\
&=\frac{\beta_0^{-\gamma'}}{\gamma'-1}\biggl(\log\frac{1+\nu}{c}\biggr)^{-\gamma'+1}\left[1-\frac{\beta_1(\gamma'-1)}{\beta_0^2}\left(\log(\frac{1+\nu}{c})\right)^{-1}\log\log(\frac{1+\nu}{c})\right]\nonumber\\
&\sim \frac{1}{\beta_0(\gamma'-1)}\left(\beta_0\log\frac{1+\nu}{c}\right)^{-\gamma'+1}\left[1-\frac{\beta_1}{\beta_0^2}\left(\log(\frac{1+\nu}{c})\right)^{-1}\log\log(\frac{1+\nu}{c})\right]^{\gamma'-1}\nonumber\\
&\sim \Biggl(\frac{1}{\beta_0\log \nu}\biggl(1-\frac{\beta_1}{\beta_0^2}\frac{\log\log\nu}{\log\nu}\biggr)\Biggr)^{\gamma'-1}
\end{align}
Thus the proof of the existence of the asymptotic solution is complete. Uniqueness follows by the Fredholm alternative. \par
We prove now the asymptotic theorem in the coordinate representation. The coordinate representation is the most convenient to get actual proofs of the $RG$ estimates, since in this representation for $x \neq 0$ 
contact terms do not occur, in such a way that composite operators are multiplicatively renormalizable. \par
In fact, the estimates in the momentum representation based on the Callan-Symanzik equations of subsect.(2.4) are rather naive, since they assume multiplicative renormalizability in the momentum representation, that is technically false.
However, the following proof of the asymptotic theorem in the coordinate representation implies also that the naive $RG$ estimate for $C_0$\footnote{And mutatis mutandis for $C_1$.} in the momentum representation,
after subtracting the contact terms, is in fact correct but for $\gamma'=0,1$. \par
To show this, we proceed writing the ansatz for the propagator in the coordinate representation, expressing the free propagator in terms of the modified Bessel function $K_1$:
\bea
&& \langle \mathcal{O}(x) \mathcal{O}(0) \rangle_{conn} \nonumber \\
&&=  \sum_{n=1}^{\infty}  \frac{1}{(2 \pi)^4} \int \frac{ R_n m^{2D-4}_n \rho^{-1}(m_n^2) } {p^2+m^2_n }\,e^{ ip\cdot x}d^4p   \nonumber \\
&&= \frac{1}{4 \pi^2 x^2} \sum_{n=1}^{\infty}    R_n m^{2D-4}_n \rho^{-1}(m_n^2) \sqrt{x^2 m^2_n} K_1( \sqrt{x^2 m^2_n} ) 
\eea
Approximating the sum by the integral using the Euler-MacLaurin formula \cite{migdal:meromorphization}, we get asymptotically:
\bea
&& \langle \mathcal{O}(x) \mathcal{O}(0) \rangle_{conn} \nonumber \\
&& \sim \frac{1}{4 \pi^2 x^2} \int_{1}^{\infty}    R_n m^{2D-4}_n \rho^{-1}(m_n^2) \sqrt{x^2 m^2_n} K_1( \sqrt{x^2 m^2_n} ) dn \nonumber \\
&& = \frac{1}{4 \pi^2 x^2} \int_{m^2_1}^{\infty}    R(m) m^{2D-4}  \sqrt{x^2 m^2} K_1( \sqrt{x^2 m^2} ) dm^2
\eea
We introduce now the dimensionless variable $z^2=x^2 m^2$:
\bea \label{R}
&& \langle \mathcal{O}(x) \mathcal{O}(0) \rangle_{conn} \nonumber \\
&& \sim \frac{1}{4 \pi^2 x^2} \int_{m^2_1}^{\infty}    R(m) m^{2D-4}  \sqrt{x^2 m^2} K_1( \sqrt{x^2 m^2} ) dm^2 \nonumber \\
&& = \frac{1}{4 \pi^2 x^2} \int_{m^2_1 x^2}^{\infty}    R(\frac{z}{x}) (\frac{z^2}{x^2})^{D-2}  z K_1(z) \frac{dz^2}{x^2} \nonumber \\
&& = \frac{1}{4 \pi^2 (x^2)^D} \int_{m^2_1 x^2}^{\infty}    R(\frac{z}{x}) z^{2D-3}   K_1(z) dz^2
\eea
In the coordinate representation the solution of the Callan-Symanzik equation (see subsect.(2.4)) is:
\begin{equation}\label{eqn:pert_general_behavior_x}
\braket{\mathcal{O}(x)\mathcal{O}(0)}_{\mathit{conn}}=
\frac{1}{(x^2)^D} \,\mathcal{G}_{0}(g(x))\, Z^2_{\mathcal{O}}(x \mu ,g(x))
\end{equation}
with the truly $RG$-invariant function $\mathcal{G}_{0}(g(x))$ admitting the expansion:
\bea
\mathcal{G}_{0}(g(x))= const(1+ \cdots)
\eea
since the correlator in the coordinate representation must be exactly conformal at the lowest non-trivial order.
Hence within the universal asymptotic accuracy:
\bea
\braket{\mathcal{O}(x)\mathcal{O}(0)}_{\mathit{conn}}
\sim \frac{1}{(x^2)^D}  \Biggl(\frac{1}{\beta_0\log(\frac{1}{x^2 \Lambda^2_{QCD}})}\biggl(1-\frac{\beta_1}{\beta_0^2}\frac{\log\log(\frac{1}{x^2 \Lambda^2_{QCD}})}{\log(\frac{1}{x^2 \Lambda^2_{QCD}})}\biggr)\Biggr)^{\frac{\gamma_0}{\beta_0}}
\eea
It follows from Eq.(\ref{R}) that it must hold:
\bea
\int_{m_1^2 x^2}^{\infty}    R(\frac{z}{x}) z^{2D-3}   K_1(z) dz^2 \sim \Biggl(\frac{1}{\beta_0\log(\frac{1}{x^2 \Lambda^2_{QCD}})}\biggl(1-\frac{\beta_1}{\beta_0^2}\frac{\log\log(\frac{1}{x^2 \Lambda^2_{QCD}})}{\log(\frac{1}{x^2 \Lambda^2_{QCD}})}\biggr)\Biggr)^{\frac{\gamma_0}{\beta_0}}
\eea
The asymptotic solution is:
\bea
R(\frac{z_0}{x}) \sim \Biggl(\frac{1}{\beta_0\log(\frac{z_0^2}{x^2 \Lambda^2_{QCD}})}\biggl(1-\frac{\beta_1}{\beta_0^2}\frac{\log\log(\frac{z_0^2}{x^2 \Lambda^2_{QCD}})}{\log(\frac{z_0^2}{x^2 \Lambda^2_{QCD}})}\biggr)\Biggr)^{\frac{\gamma_0}{\beta_0}}
\eea
Indeed, the universal part of $R(\frac{z}{x})$ is actually $z$ independent and therefore we can put it, for any fixed $z=z_0$, outside the integral over $z$ in the limit $x \rightarrow 0$:
\bea
&&\int_{m_1^2 x^2}^{\infty}    R(\frac{z}{x}) z^{2D-3}   K_1(z) dz^2 \sim R(\frac{z_0}{x}) \int_{0}^{\infty}    z^{2D-3}   K_1(z) dz^2 \nonumber \\
&&\sim \Biggl(\frac{1}{\beta_0\log(\frac{1}{x^2 \Lambda^2_{QCD}})}\biggl(1-\frac{\beta_1}{\beta_0^2}\frac{\log\log(\frac{1}{x^2 \Lambda^2_{QCD}})}{\log(\frac{1}{x^2 \Lambda^2_{QCD}})}\biggr)\Biggr)^{\frac{\gamma_0}{\beta_0}}
\eea
since the integral:
\bea
\int_{0}^{\infty}    z^{2D-3}   K_1(z) dz^2 
\eea
is convergent for $D>1$ because $K_1$ has a simple pole in $z=0$ and decays exponentially for large $z$. Therefore, within the universal asymptotic accuracy:
\bea
R(\frac{z_0}{x}) \sim  Z^2_{\mathcal{O}}(x \mu ,g(x))
\eea
and the naive $RG$ estimate in momentum space is in fact correct but for $\gamma'=0,1$. \par
Indeed, we have just proved that the universal part of the residues $R_n$ determined by the integral equations in the coordinate representation and in the momentum representation is the same. Since in the coordinate representation the $RG$ estimate
is certainly correct because of the lack of contact terms, it follows that the asymptotic behavior in the momentum representation is computable using the sum of free propagators with the residues determined by the coordinate representation as input.
But then, after \emph{subtracting} the contact terms that arise in the sum of free propagators, the asymptotic behavior in the momentum representation is \emph{computed} by the integral in Eq.(\ref{eqn:esp_ntl}), that coincides with the naive $RG$ estimate of subsect.(2.4) \cite{MB} but for $\gamma'=0,1$. \par 
The extension to any integer spin $s$ is an easy corollary. It is only necessary to prove that:
\bea
&&\sum_{n=1}^{\infty}  P^{(s)} \big(\frac{p_{\alpha}}{m^{(s)}_n}\big) \frac{m^{(s)2D-4}_n Z_n^{(s)2}  \rho_s^{-1}(m^{(s)2}_n)}{p^2+m^{(s)2}_n  } \nonumber \\
&&=P^{(s)} \big(\frac{p_{\alpha}}{p} \big)  \, p^{2D-4}   \sum_{n=1}^{\infty} \frac{Z_n^{(s)2}   \rho_s^{-1}(m^{(s)2}_n)  }{p^2+m^{(s)2}_n  }
     + \cdots
\eea
where the dots represent contact terms and $P^{(s)} \big(\frac{p_{\alpha}}{p} \big)$ is the projector obtained substituting $-p^2$  to $m_n^2$ in  $P^{(s)} \big(\frac{p_{\alpha}}{m_n} \big)$. The proof is as follows.
$m^{(s)2D-4}_n P^{(s)} \big(\frac{p_{\alpha}}{m^{(s)}_n}\big)$ is a polynomial in powers of $m^2_n$. To each monomial $m^{2d}_n$ occurring in this polynomial we can substitute either $p^{2d}$ or $- p^{2d}$, for $d$ even or $d$ odd respectively,
up to contact terms, because of Eq.(\ref{b0}) and Eq.(\ref{b2}). This is the same as substituting  $-p^2$  to $m_n^2$ in  $P^{(s)} \big(\frac{p_{\alpha}}{m_n} \big)$ since for $d$ even we always get a positive sign.
The asymptotic theorem for any spin follows. \par
For completeness we write explicitly the spin-$1$ and the spin-$2$ propagators as determined by the asymptotic theorem. We employ Veltman conventions for Euclidean and Minkowski propagators  (see Appendix F in \cite{Velt2}). \par
For spin $1$:
\bea
&& \int \langle \mathcal{O}^{(1)}_{\alpha}(x) \mathcal{O}^{(1)}_{\beta}(0) \rangle_{conn}\,e^{-ip\cdot x}d^4x  \nonumber \\
&& \sim \sum_{n=1}^{\infty} (\delta_{\alpha \beta} +  \frac{p_{\alpha} p_{\beta}}{m^{(1) 2}_n}) \frac{m^{(1)2D-4}_n Z_n^{(1)2}  \rho_1^{-1}(m^{(1)2}_n)}{p^2+m^{(1)2}_n  } \nonumber \\
&& \sim  \, p^{2D-4}     (\delta_{\alpha \beta} -  \frac{p_{\alpha} p_{\beta}}{p^2}) \sum_{n=1}^{\infty}   \frac{Z_n^{(1)2}   \rho_1^{-1}(m^{(1)2}_n)  }{p^2+m^{(1)2}_n  }
     + \cdots
\eea
For spin 2:
\bea
&& \int \langle \mathcal{O}^{(2)}_{\alpha \beta}(x) \mathcal{O}^{(2)}_{\gamma \delta}(0) \rangle_{conn}\,e^{-ip\cdot x}d^4x  \nonumber \\
&& \sim \sum_{n=1}^{\infty} \bigg[\frac{1}{2} \eta_{\alpha \gamma}(m^{(2)}_n) \eta_{\beta \delta}(m^{(2)}_n) + \frac{1}{2} \eta_{\beta \gamma}(m^{(2)}_n)  \eta_{\alpha \delta}(m^{(2)}_n) - \frac{1}{3} \eta_{\alpha \beta}(m^{(2)}_n)  \eta_{\gamma \delta}(m^{(2)}_n) \bigg]  \frac{m^{(2)2D-4}_n Z_n^{(2)2}  \rho_2^{-1}(m^{(2)2}_n)}{p^2+m^{(2)2}_n  } \nonumber \\
&&\sim  \, p^{2D-4}  \bigg[\frac{1}{2}  \eta_{\alpha \gamma}(p) \eta_{\beta \delta}(p) + \frac{1}{2} \eta_{\beta \gamma}(p)  \eta_{\alpha \delta}(p) - \frac{1}{3} \eta_{\alpha \beta}(p)  \eta_{\gamma \delta}(p) \bigg]     \sum_{n=1}^{\infty}   \frac{Z_n^{(2)2}   \rho_2^{-1}(m^{(2)2}_n)  }{p^2+m^{(2)2}_n  }+\cdots
\eea
where:
\bea
\eta_{\alpha \beta}(m)= \delta_{\alpha \beta} +  \frac{p_{\alpha} p_{\beta}}{m^2}
\eea
and:
\bea
\eta_{\alpha \beta}(p)= \delta_{\alpha \beta} -  \frac{p_{\alpha} p_{\beta}}{p^2}
\eea
Some observations are in order. \par
Each massive propagator is conserved only on the respective mass shell. However, after subtracting the sum of contact terms denoted by the dots (a polynomial of \emph{finite} degree in the momentum representation with diverging coefficients), the resulting massless projector implies off-shell conservation, as if the large-$N$ $QCD$ propagators
were saturated by massless particles only. This is necessary to match $QCD$ perturbation theory (with massless quarks). For a direct check see \cite{chetyrkin:TF,chet:tensore}. \par
In the spin-$2$ case the massless projector contains a factor of $\frac{1}{3}$ in the last term, that descends from the massive case, and not of $\frac{1}{2}$, that would occur for a truly physical massless spin-$2$ particle according to van Dam-Veltman-Zakharov  discontinuity \cite{Velt,Zak}. This factor of $\frac{1}{3}$ occurs also in perturbative computations of the correlator of the stress-energy tensor in $QCD$ \cite{chet:tensore}. Indeed,
a spin-$2$ glueball in $QCD$ is not a graviton.

 \thispagestyle{empty}



\begin{thebibliography}{99}
\bibitem{MB} M. Bochicchio, S. P. Muscinelli, \emph{Ultraviolet asymptotics of glueball propagators}, \href{http://arxiv.org/abs/arXiv:1304.6409}{hep-th/1304.6409}. 
\bibitem{chet:tensore}
M. F. Zoller, K. G. Chetyrkin, \emph{OPE of the energy-momentum
tensor correlator in massless QCD}, \href{http://arxiv.org/abs/1209.1516}{hep-ph/1209.1516}.
\bibitem{chetyrkin:scalar}
K.G. Chetyrkin, B.A. Kniehl, and M. Steinhauser, \emph{Hadronic Higgs Boson Decay to Order $\alpha^4$}, \emph{Phys. Rev. Lett.} {\bf 79} 353 (1997) \href{http://arxiv.org/abs/hep-ph/9705240}{[hep-ph/9705240]}.
\bibitem{chetyrkin:pseudoscalar}
K.G. Chetyrkin, B.A. Kniehl, M. Steinhauser, W.A. Bardeen, \emph{Effective QCD Interactions of CP-odd Higgs Bosons at
Three Loops}, \emph{Nucl. Phys. B} {\bf 535} 3 (1998) \href{http://arxiv.org/abs/hep-ph/9807241}{[hep-ph/9807241]}.
\bibitem{Kataev:1981gr}
 A. L. Kataev, N. V. Krasnikov, A. A. Pivovarov,
 \emph{Two Loop Calculations For The Propagators Of Gluonic Currents}, \emph{Nucl. Phys. B} {\bf 198} (1982) 508, 
  [Erratum-ibid. {\bf 490} (1997) 505] \href{http://arxiv.org/abs/hep-ph/9612326}{[hep-ph/9612326]}.
\bibitem{ferretti:new_struct}
G. Ferretti, R. Heise, K. Zarembo, \emph{New integrable structures in large-N QCD}, \emph{Phys. Rev. D} \textbf{70} (2004) 074024 \href{http://arxiv.org/abs/hep-th/0404187}{[hep-th/0404187]}.
\bibitem{boch:quasi_pbs}
M. Bochicchio, \emph{Quasi BPS Wilson loops, localization of loop
equation by homology and exact beta function in the
large-N limit of $SU(N)$ Yang-Mills theory}, JHEP \textbf{0905} 116 (2009) \href{http://arxiv.org/abs/0809.4662}{[hep-th/0809.4662]}.
\bibitem{MB0} 
M. Bochicchio, \emph{Exact beta function and glueball spectrum in large-N Yang Mills theory}, \emph{PoS EPS-HEP2009: {\bf 075}(2009)} \href{http://arxiv.org/abs/0910.0776}{[hep-th/0910.0776]}.
\bibitem{boch:crit_points}
M. Bochicchio, \emph{Glueballs in large-N $YM$ by localization on critical points}, \href{http://arxiv.org/abs/1107.4320}{hep-th/1107.4320}, extended version of the talk at the Galileo Galilei Institute
Conference "Large-$N$ Gauge Theories", Florence, Italy, May 2011.
\bibitem{boch:glueball_prop}
M. Bochicchio, \emph{Glueball propagators in large-N YM}, \href{http://arxiv.org/abs/1111.6073}{hep-th/1111.6073}. 
\bibitem{Top} M. Bochicchio, \emph {Yang-Mills mass gap at large-N, topological quantum field theory and hyperfiniteness}, \href{http://arxiv.org/abs/1202.4476}{hep-th/1202.4476},
a byproduct of the Simons Center workshop "Mathematical Foundations of Quantum Field Theory", Stony Brook, USA, Jan 16-20 (2012).
\bibitem{Me1} H. B. Meyer, M. J. Teper, \emph{Glueball Regge trajectories and the Pomeron -- a lattice study --}, \emph{Phys.Lett. B} {\bf  605} 344 (2005) \href{http://arxiv.org/abs/hep-ph/0409183}{[hep-ph/0409183]}.
\bibitem{Me2} H. B. Meyer, \emph{Glueball Reggge Trajectories}, \href{http://arxiv.org/abs/arXiv:hep-lat/0508002}{[hep-lat/0508002]}.
\bibitem{L1} B. Lucini, M. Teper, U. Wenger, \emph{Glueballs and k-strings in SU(N) gauge theories: calculations with improved operators}, JHEP \textbf{0406} 012 (2004) \href{http://arxiv.org/abs/hep-lat/0404008}{[hep-lat/0404008]}.
\bibitem{L2} B. Lucini, A. Rago, E. Rinaldi, \emph{Glueball masses in the large N limit}, JHEP {\bf 1008} 119 (2010)   \href{http://arxiv.org/abs/arXiv:1007.3879}{[hep-lat/1007.3879]}. 
\bibitem{Mal} O. Aharony, S. S. Gubser, J. Maldacena, H. Ooguri, Y. Oz,  \emph{Large $N$Field Theories, String Theory and Gravity},
Phys. Rept. {\bf 323} 183 (2000) \href{http://arxiv.org/abs/hep-th/9905111}{[hep-th/9905111]}.
\bibitem{Witten} E. Witten, \emph{Anti-de Sitter Space, Thermal Phase Transition, And Confinement In Gauge Theories}, \emph{Adv. Theor. Math. Phys.} {\bf2} 505 (1998) \href{http://arxiv.org/abs/hep-th/9803131}{[hep-th/9803131]}.
\bibitem{KS1}  I. R. Klebanov, M. J. Strassler, \emph{Supergravity and a Confining Gauge Theory: Duality Cascades and $\chi$SB-Resolution of Naked Singularities},
JHEP {\bf 0008} 052 (2000) \href{http://arxiv.org/abs/hep-th/0007191}{[hep-th/0007191]}.
\bibitem{KS2}   M. J. Strassler, \emph{The Duality Cascade}, \href{http://arxiv.org/abs/hep-th/0505153}{hep-th/0505153}.
\bibitem{PS}  J. Polchinski, M. J. Strassler, \emph{Hard scattering and gauge/string duality}, \emph{Phys. Rev. Lett.} {\bf 88} (2002) 031601 \href{http://arxiv.org/abs/hep-th/0109174}{[hep-th/0109174]}.
\bibitem{PS2} R. C. Brower, J. Polchinski, M. J. Strassler, C.-I. Tan, \emph{The Pomeron
and Gauge/String Duality}, JHEP {\bf 0712} (2007) 005 \href{http://arxiv.org/abs/hep-th/0603115}{[hep-th/0603115]}.
\bibitem{Softwall} A. Karch, E. Katz, D. T. Son, M. A. Stephanov, \emph{Linear Confinement and AdS/QCD}, \emph{Phys. Rev. D} {\bf 74} 015005 (2006) \href{http://arxiv.org/abs/hep-ph/0602229}{[hep-ph/0602229]}.
\bibitem{Brower2}  R. C. Brower, M. Djuric, C.-I Tan, \emph{Odderon in Gauge/String Duality} JHEP {\bf 0907} 063 (2009) \href{http://arxiv.org/abs/arXiv:0812.0354}{[hep-th/0812.0354]}.
\bibitem{Brower1} R. C. Brower, S. D. Mathur,  \emph{Glueball Spectrum for QCD from AdS Supergravity Duality}, \emph{Nucl. Phys. B} {\bf 587} 249 (2000) \href{http://arxiv.org/abs/hep-th/0003115}{[hep-th/0003115]}.
\bibitem{forkel} H. Forkel, \emph{Holographic glueball structure}, \emph{Phys. Rev. D} {\bf 78} 025001 (2008) \href{http://arxiv.org/abs/0711.1179}{[hep-ph/0711.1179]}.
\bibitem{italiani}  P. Colangelo, F. de Fazio, F. Jugeau, S. Nicotri, \emph{Investigating AdS/QCD duality through scalar glueball correlators}, \emph{ Int. J. Mod. Phys. A} {\bf 24} 4177 (2009)
\href{http://arxiv.org/abs/0711.4747}{[hep-ph/0711.4747]}.
\bibitem{forkel:holograms} H. Forkel, \emph{Glueball correlators as holograms}, \href{http://arxiv.org/abs/0808.0304}{hep-ph/0808.0304}.
\bibitem{forkel:ads_qcd}
H. Forkel, \emph{AdS/QCD at the correlator level}, \emph{PoS(Confinement8)} {\bf 184} \href{http://arxiv.org/abs/0812.3881}{[hep-ph/0812.3881]}.
\bibitem{krasnitz:cascading2}
M. Krasnitz,  
\emph{A two point function in a cascading $\mathcal{N} =1$
gauge theory from supergravity}, \href{http://arxiv.org/abs/hep-th/0011179}{hep-th/0011179}.
\bibitem{krasnitz:cascading}
M. Krasnitz,
 \emph{Correlation functions in a cascading $\mathcal{N} =1$ gauge
theory}, JHEP {\bf 0212} (2002) 048 \href{http://arxiv.org/abs/hep-th/0209163}{[hep-th/0209163]}.
\bibitem{KSh} I. I. Kogan, M. Shifman, \emph{Two Phases of Supersymmetric Gluodynamics}, \emph{Phys. Rev. Lett.}{ \bf 75} 2085 (1995) \href{http://arxiv.org/abs/hep-th/9504141}{[hep-th/9504141]}.
\bibitem{ML} B. Lucini, M. Panero, \emph{SU(N) gauge theories at large N}, \emph{Physics Reports} {\bf 526} (2013) 93 \href{http://arxiv.org/abs/1210.4997}{[hep-th/1210.4997]}.
\bibitem{Braga1} H. Boschi-Filho, N. R. F. Braga, \emph{QCD/String holographic mapping and glueball mass spectrum}, \emph{Eur. Phys. J. C} {\bf 32} (2004) 529 \href{http://arxiv.org/abs/hep-th/0209080}{[hep-th/0209080]}.
\bibitem{Braga2} H. Boschi-Filho, N. R. F. Braga, H. L. Carrion, \emph{Glueball Regge trajectories from gauge/string duality and the Pomeron}, \emph{Phys. Rev. D} {\bf 73}(2006) 0407901 \href{http://arxiv.org/abs/hep-th/0507063}{[hep-th/0507063]}.
 \bibitem{'t hooft:large_n}
G. 't Hooft, \emph{Nucl. Phys. B} \textbf{ 72} 461 (1974).
\bibitem{migdal:multicolor}
A. Migdal, \emph{Multicolor QCD as a dual-resonance theory}, \emph{Annals of Physics} \textbf{109} 365 (1977).
\bibitem{ZI} C. Itzykson, J-B. Zuber, \emph{Quantum Field Theory}, McGraw-Hill.
\bibitem{polyakov:gauge}
A. M. Polyakov, \emph{Gauge Fields and Strings}, Harwood Academic Publishers.
\bibitem{Velt} H. Van Dam, M. Veltman, \emph{Massive and Mass-Less Yang-Mills And Gravitational Fields}, \emph{Nucl. Phys. B} {\bf 22} 397 (1970).
\bibitem{Zak} V. I. Zakharov, JETP Lett. {\bf12} (1970) 312 [Pisma Zh. Eksp. Teor. Fiz. {\bf 12} (1970) 447].
\bibitem{Francia1} D. Francia, J. Mourad, A. Sagnotti, \emph{Current Exchanges and Unconstrained Higher Spins}, \emph{Nucl. Phys. B} {\bf 773} (2007) 203 \href{http://arxiv.org/abs/hep-th/0701163}{[hep-th/0701163]}.
\bibitem{Francia2} D. Francia, \emph{Geometric massive higher spins and current exchanges}, \emph{Fortsh. Phys.} {\bf 56} (2008) 800	\href{http://arxiv.org/abs/arXiv:0804.2857}{[hep-th/0804.2857]}.
\bibitem{chetyrkin:TF}
K. G. Chetyrkin, A. Maier, \emph{Massless correlators of vector, scalar and tensor currents in position space at orders $\alpha_s^3$ and $\alpha_s^4$: explicit analytical results}, \emph{Nucl. Phys. B} {\bf 844} 266 (2011) \href{http://arxiv.org/abs/1010.1145}{[hep-ph/1010.1145]}.
\bibitem{1}  V. A. Novikov, M. A. Shifman, A. I. Vainshtein, V. I. Zakharov, \emph{Nucl. Phys. B} {\bf 165} (1980) 67.
\bibitem{2} V. A. Novikov, M. A. Shifman, A. I. Vainshtein, V. I. Zakharov, \emph{Nucl. Phys. B} {\bf 191} (1981) 301.
\bibitem{migdal:meromorphization}
A. Migdal, \emph{Meromorphization of Large N QFT}, \href{http://arxiv.org/abs/1109.1623}{hep-th/1109.1623}.
\bibitem{Velt2} M. Veltmann, \emph{Diagrammatica}, Cambridge University Press.
\bibitem{A1}
J. Mondejar, A. Pineda, \emph{Constraints on Regge models from perturbation theory},   JHEP {\bf 0710} (2007) 061 \href{http://arxiv.org/abs/0704.1417}{[hep-th/0704.1417]}.
\bibitem{A2}
J. Mondejar, A. Pineda, \emph{$1/N_c$ and 1/n preasymptotic corrections to Current-Current correlators}, JHEP {\bf 0806} (2008) 039 \href{http://arxiv.org/abs/0803.3625}{[hep-th/0803.3625]}.
\end{thebibliography}
\end{document}